\documentclass[twocolumn]{aastex62}

\usepackage{epsfig}
\usepackage{ifthen}
\usepackage[caption=false]{subfig}
\usepackage{natbib}
\usepackage{amssymb, amsmath}
\usepackage{appendix}

\usepackage{siunitx}
\usepackage{epstopdf}
\usepackage[T1]{fontenc}
\usepackage{booktabs}
\usepackage{multirow}
\usepackage{paralist}
\usepackage{rotating} 
\usepackage{upgreek}
\newcolumntype{L}[1]{>{\raggedright\arraybackslash}p{#1}}
\newcolumntype{C}[1]{>{\centering\arraybackslash}p{#1}}
\newcolumntype{R}[1]{>{\raggedleft\arraybackslash}p{#1}}
\graphicspath{{./figures/}}
\usepackage[utf8]{inputenc}
\graphicspath{{./}{figures/}}

\received{June 02, 2022}
\revised{June 29, 2022}
\accepted{July 15, 2022}
\submitjournal{ApJS}

\shorttitle{iCOMs into HMSFRs}
\shortauthors{O.S. Rojas-García et al.}

\begin{document}

\title{Interstellar Complex Organic Molecules in SiO-traced massive outflows }


\correspondingauthor{O. S. Rojas-García}
\email{sergio.rojas@inaoep.mx}

\author{O. S. Rojas-García}
\affiliation{Instituto Nacional de Astrofísica, Óptica y Electrónica, Luis E. Erro 1, Tonantzintla, Puebla, C.P. 72840, México}

\author{A. I. Gómez-Ruiz}
\affiliation{Instituto Nacional de Astrofísica, Óptica y Electrónica, Luis E. Erro 1, Tonantzintla, Puebla, C.P. 72840, México}

\author{A. Palau}
\affiliation{Instituto de Radioastronomía y Astrofísica, Universidad Nacional Autónoma de México, Antigua Carretera a Pátzcuaro \# 8701,
Ex-Hda. San José de la Huerta, Morelia, Michoacán, México C.P. 58089}

\author{M. T. Orozco-Aguilera}
\affiliation{Instituto Nacional de Astrofísica, Óptica y Electrónica, Luis E. Erro 1, Tonantzintla, Puebla, C.P. 72840, México}

\author{M. Chavez Dagostino}
\affiliation{Instituto Nacional de Astrofísica, Óptica y Electrónica, Luis E. Erro 1, Tonantzintla, Puebla, C.P. 72840, México}

\author{S. E. Kurtz}
\affiliation{Instituto de Radioastronomía y Astrofísica, Universidad Nacional Autónoma de México, Antigua Carretera a Pátzcuaro \# 8701,
Ex-Hda. San José de la Huerta, Morelia, Michoacán, México C.P. 58089}

\nocollaboration
\begin{abstract}
The interstellar medium contains dust and gas, in which at high densities and cold conditions molecules can proliferate. Interstellar Complex Organic Molecules (iCOMs) are C-bearing species that contain at least six atoms. As they are detected in young stellar objects, iCOMs are expected to inhabit early stages of the star formation evolution. In this study, we try to determine which iCOMs are present in the outflow component of massive protostars. To do this, we analyzed the morphological extension of blue- and red-shifted iCOMs emission in a sample of eleven massive protostars employing mapping observations at one mm within a $\sim$1 GHz bandwidth for both IRAM-30m and APEX telescopes. We modeled the iCOMs emission of the central pointing spectra of our objects using the XCLASS LTE radiative transfer code. We detected the presence of several iCOMs such as: CH$_3$OH, $^{13}$CH$_3$OH, CH$_3$OCHO, C$_2$H$_5$C$^{15}$N, and ($c-$C$_3$H$_2$)CH$_2$. In G034.41+0.24, G327.29-0.58, G328.81+0.63, G333.13-0.43, G340.97-1.02, G351.45+0.66 and G351.77-0.54, the iCOMs lines show a faint broad line profile. Due to the offset peak positions of the blue- and red-shifted emission, covering from $\sim$ 0.1 to 0.5 pc, these wings are possibly related to movements external to the compact core, such as large-scale low-velocity outflows. We have also established a correlation between the parent iCOM molecule CH$_3$OH and the shock tracer SiO, reinforcing the hypothesis that shock environments provide the conditions to boost the formation of iCOMs via gas-phase reactions.





\end{abstract}



\section{Introduction} \label{sec:intro}
Interstellar clouds, star formation regions,  circumstellar envelopes of AGB stars, and planetary nebulae, have proven to hold the physical conditions to harbor a rich molecular content. Among them,  
many of these molecules are organic in nature. This fact is especially remarkable in the larger molecules, of which those with at least six atoms have been defined as iCOMs (\textit{Interestellar Complex Organic Molecules})\citep{Herbst2009, Ceccarelli2017ApJ...850..176C}.

Recent efforts have been conducted to establish the mechanism to explain the abundances of iCOMs in several interstellar environments. The dominant scenarios are the dust grain chemistry and the warm gas-phase reactions \citep{Garrod2006A&A...457..927G, Taquet_2012}. These two scenarios could work simultaneously, and are thought to take place as follows. In a first cold pre-stellar phase, atoms and small molecules are stuck into dust grains and are subsequently hydrogenated and oxidised, forming mainly ice mantles of H$_2$O, CO$_2$, NH$_3$, and CH$_4$. In a subsequent phase, triggered by the temperature increase due to the star formation process or by non-thermal processes, the mantle components are released to the gas phase where additional reactions form the iCOMs. Alternatively, there is also the possibility that the increase in temperature triggers further reactions within processed (irradiated by UV photons or cosmic-rays) grain mantles, forming the iCOMs in the dust grain surfaces, which will be later desorbed (see the review by \citet{Ceccarelli22} for further details).

The shock waves produced by outflow movements during the star-formation process are one of the environments that can facilitate chemical reactions leading to more complex chemistry.
This phenomenon is present during the early phases of the formation of low- and high-mass stars \citep{arce}. 

The supersonic material ejected from the inner regions of the protostellar system generates shock waves along its path, causing an increase of density and temperature in the surrounding material. Such conditions could boost chemical reactions that otherwise wouldn't take place in the cold and quiescent ISM.  

The increase in the abundances of certain molecules within the shocked regions of low-mass outflows have unveiled a particular shock-induced chemistry, the prototype of which has been the low-mass protostar L1157-mm \citep{bachiller2001}. Other examples are: BHR71 \citep[][low mass]{bourke1997flujo-molecular-activo}, HH 114-MMS \citep[][low mass]{tafa2013active}, IRAS 16547-4247 \citep[][high mass]{Garay2007A&Aoutflow,DeSimone2020A&A...640A..75D} and  IRAS 17233-3606 \citep[][high mass]{leurini2011cinematica17233-3606, leurini2013evidencia17233-3606, leurini2013gas}.

The extensive studies in the archetype outflow L1157-mm have revealed the rich chemistry induced by shocks. One of the remarkable discoveries was the detection of iCOMs within these shocked regions \citep{Bachiller95}, which suggests a relation between the shocks and iCOMs formation \citep[i.e.][]{leflo2017}. 

In the high-mass regime, a few outflows have also demonstrated a peculiar chemical richness, however, an unambiguous analysis of the high-velocity emission has been difficult given the outflow multiplicity usually found in these environments, together with the usual low-angular resolution observations. This leads to the confusion of the different spectral components that may not belong to the same outflow \citep[i.e.][]{leurini2013evidencia17233-3606}. A recent systematic study has been presented in the intermediate- to high-mass protostellar object IRAS 20126$+$4104 \citep[][]{palau2017}, indicating that iCOMs in this source may arise from the disc and the dense/hot region along with the outflow. The abundance of the iCOMs shows an enhancement at the outflow positions. Such enhancement has also been related to shocks in the accretion disc of G328.2551-0.5321, a presumable pre-hot core source \citep{Csengeri2019A&A}. 

To improve our knowledge of the presence of iCOMs in the outflow phase, we study a set of outflow candidates using mapping observations from single-dish telescopes. Employing the integrated maps over the line wing emission, we tried to isolate the molecular emission of iCOMs originated by the shock itself. The presence of outflow structures in our sample was confirmed by their association with high-velocity wings of SiO emission (Rojas-García et, al. (in prep)).

\section{Observations} \label{sec:sample}
\subsection{Sample selection}
Our data set consists of eleven massive protostars associated with Extended Green Objects (EGOs), and APEX Telescope Large Area Survey of the Galaxy (ATLASGAL), sources \citep{Cyganowski2008, Csengeri2014A&A...565A..75C}, three of them (the northern ones) belong to the EGO survey, whereas eight of them (the southern ones) come from the ATLASGAL survey. These specific sources were selected based on a previous survey focused on the detection of strong emitters of SiO 2-1 and 1-0 transitions\footnote{This investigation will be presented in a separate paper, but the results of this survey are already published in a master's thesis that can be accessed in the following link: \url{https://inaoe.repositorioinstitucional.mx/jspui/handle/1009/851}}. As it is well known, the SiO is a selective tracer of shocked environments, among these, the outflow phase of the star formation process \citep{Tafalla10, tafa2011, duarte2014, Csengeri2016}. 
\subsection{Data}
Our sample were observed by means of mapping observations covering a 100$''$ square area in each high mass star forming region (HMSFR). Eight of the sources belong to the southern sky and were observed by APEX telescope with a spectral coverage from  216.681 to 217.680 ($\Delta v = 0.34$ kms$^{-1}$) at $\theta_{\text{HPBW}} \sim 30.5''$ with a positional accuracy of $4''$. These observations were taken in two observation sessions, the first one during May 21, 22, 24, and 26, and the second during September 22 and 26 of 2010.

The remaining three sources belong to the northern sky and were observed with the IRAM-30m telescope. In this case, the spectral coverage ranges from 216.925 to 217.791 GHz ($\Delta v = 2.76$ kms$^{-1}$) at $\theta_{\text{HPBW}} \sim 11.94''$ with a positional accuracy of $2''$. These observations were taken on February and March 19 and 21 of 2011 (Rojas-García et, al. in prep). 

The complete list of these objects is shown in Table \ref{sample}.  

\begin{table}[!htpb]
	\centering
	\caption{List of HMSFRs mapped within $\sim 1$ GHz spectral coverage at 1 mm. Sources at positive declinations were observed with IRAM whereas at negative declinations, sources were observed with APEX. The distances were obtained from \citet{he2012,Wienen12,miettinen2006}.}
	\begin{tabular}{ccccccccccccccccc}
		\toprule
		GLIMPSE  &\multicolumn{2}{c}{Coordinates} &  \multirow{2}*{Distance}\\
		\cmidrule(r){2-3}
		sources & $\alpha$ (J2000)& $\delta$  (J2000)\\
		\midrule
G034.26+0.15     &	18:53:16.40 &	01:15:07.00 	 &3.64$\pm$0.36\\
G034.41+0.24     &	18:53:17.90 &	01:25:25.00 	 &3.59$\pm$0.36\\
G035.13$-$0.74   &	18:58:06.40 &	01:37:01.00 	 &2.34$\pm$0.38\\
G327.29$-$0.58   &	15:53:08.56 &	-54:37:05.50	 &3.03$\pm$0.65\\
G328.81+0.63     &	15:55:48.56 &	-52:43:08.40	 &2.83$\pm$0.65\\
G333.13$-$0.43   &	16:21:02.67 &	-50:35:13.30	 &3.54$\pm$0.65\\
G337.92$-$0.48   &	16:41:10.72 &	-47:08:07.20	 &3.10$\pm$0.65\\
G340.97$-$1.02   &	16:54:57.30 &	-45:09:04.00	 &2.27$\pm$0.65\\
G351.45+0.66     &	17:20:54.34 &	-35:45:03.80	 &0.84$\pm$0.65\\
G351.77$-$0.54   &	17:26:42.93 &	-36:09:20.00	 &0.59$\pm$0.65\\
G353.41$-$0.36   &	17:30:26.86 &	-34:41:50.50	 &3.50$\pm$0.65\\
		\bottomrule                                                  
	\end{tabular}\label{sample}
\end{table} 

\subsection{Reduction and observational results}
The datacubes were smoothed to a nominal spectral resolution of 2.76 kms$^{-1}$ (our lower spectral resolution), to improve the signal-to-noise ratio and to make a consistent analysis throughout the sample. 
We identified and avoided the damaged observations along each datacube for all the sources. Then, a baseline was subtracted (in most of the sources the baseline was fitted with a linear curve), and the resulting spectra were converted from T$_A^*$ to T$_{mb}$ according to their corresponding efficiencies, for IRAM $n_{ff}=$ 0.915 and $n_\text{mb}=$	0.57 and for APEX $n_{ff}=$ 0.970 and $n_\text{mb}=$ 0.75.

We then analyzed the 12 strongest lines observed in most of the objects (common lines among the sample) and did a preliminary identification by visual inspection based on the coincidence of the line emission peak frequency and the nominal frequency expected for molecules in the \textit{Splatalogue} catalog. This line identification was later confirmed by means of an LTE synthetic model and, after this procedure, we obtained the molecular species listed in Table \ref{detected_molecules}.
 \begin{table*}[!htpb]
	\centering
	\caption{Full list of the identified iCOMs after the XCLASS modeling. The main properties for each detected species are displayed in columns left to right. (1) Chemical formulae, (2) molecule name, (3) rest frequency, (4) Resolved quantum number transition, (5) Einstein emission coefficient A$_{ij}$, (6) upper energy level  and (7) upper degeneracy level.}
	
	\begin{tabular}{ccccccccccccccccc}
		\toprule
 \multirow{2}*{Molecule}     &     \multirow{2}*{Name}      &      Frequency &    Transition line  &    Log10 (A$_{ij}$) &  $E_u/\kappa$	&    \multirow{2}*{$g_u$}   \\
                             &                              &                          (GHz)             &      &         & (K)             &       \\
\midrule
CH$_3$OCHO          &   Methyl Formate	        &  216.830197	&   18( 2,16)-17( 2,15) E	 &  -3.8297 	&   105.677	& 74  \\
CH$_3$OCHO       	&   Methyl Formate	        &  216.838889	&   18( 2,16)-17( 2,15) A	 &  -3.8297 	&   105.666 	& 37  \\
CH$_3$OH            &   Methanol	            &  216.945559	&   5( 1, 4)- 4( 2, 2)	     &  -4.9158 	&   55.871	& 11  \\
CH$_3$OCHO          &   Methyl Formate	        &  216.965900	&   20( 1,20)-19( 1,19) A	 &  -3.8149 	&   111.481	& 82  \\
CH$_3$OCHO          &   Methyl Formate	        &  216.967994	&   20( 1,20)-19( 0,19) E	 &  -4.6118 	&   111.498	& 82  \\
C$_2$H$_5$C$^{15}$N &   Ethyl Cyanide 	        &  217.190187	&   12(3,10)-11(2,9)  	     &  -4.5314	    &    31.953	& ... \\
CH$_3$OCHO      	&   Methyl Formate	        &  217.237558	&   7( 3, 5)- 6( 1, 6) E	 &  -6.5284 	&   22.527	& 30  \\
CH$_3$OCHO      	&   Methyl Formate	        &  217.298353	&   8(5,4)-8(3,5) E	         &  -6.7361 	&   37.837	&  34  \\
CH$_3$OCHO          &   Methyl Formate	        &  217.301326	&   7( 3, 5)- 6( 1, 6) A	 &  -6.5259 	&   22.511	& 30  \\
$^{13}$CH$_3$OH     &   Methanol     	        &  217.399550	&   10( 2, 8)- 9( 3, 7) ++	 &  -4.8153 	&   162.413	& 21  \\
(c-C$_3$H$_2$)CH$_2$&   Methylenecyclopropene	&  217.429650	&   16(4,13) - 16(2,14)	     &  -5.8861 	&   102.323	& 165  \\
\bottomrule                                                  
\end{tabular}\label{detected_molecules}
\end{table*} 

The integrated area of the lines was calculated within their Full Width at Zero Power (FWZP) widths. Their corresponding uncertainties were computed according to standard formalism: $\sigma_\text{area}=\sqrt{N}\times\Delta v \times \sigma$, where $\Delta v$ is the velocity resolution, $\sigma$ is the RMS noise level per channel, and $N$ is the number of channels within the integrated velocity range. The resulting values are reported in Table \ref{Tmb_Int}. 

\begin{table*}[!htpb]
	\centering
	\small
	\caption{Integrated area over the strongest 12 molecular emission lines detected in most of the sources. The labeled iCOMs were added after their confirmation by LTE modeling. Column (1) contains the chemical formula of the corresponding emitter. Column (2) lists their corresponding rest frequency. Columns (3) to (8) display the integrated area $\int T_{Mb}\ dv$ in [K km s$^{-1}$] over the FWZP width for all the sources (top labeled in columns 3 to 8).}  
	\begin{tabular}{cccccccccc} 
		\toprule
		 \multirow{2}*{Molecule} &  Frequency  & G327.29-0.58 & G328.81+0.63  & G333.13-0.43 & G337.92-0.48 & G340.97-1.02 & G351.45+0.66 \\                                                                                                        
    &    (GHz)    & [K km s$^{-1}$] &  [K km s$^{-1}$] &  [K km s$^{-1}$]  &  [K km s$^{-1}$]   &  [K km s$^{-1}$]  &  [K km s$^{-1}$]      \\   
    (1)& (2)& (3)& (4)& (5)& (6)& (7)& (8) \\
\midrule  
CH$_3$OCHO          &  216.830197	& 1.5$\pm$0.2  & --- & --- & ---  & --- & --- \\
CH$_3$OCHO       	&  216.838889	& 3.2$\pm$0.3  & --- & --- & ---  & --- & --- \\
CH$_3$OH            &  216.945559	& 6.7$\pm$0.6  & 3.9$\pm$0.2 & 2.0$\pm$0.1 & 2.1$\pm$0.3  & 1.5$\pm$0.4 & 4.1$\pm$0.1 \\
CH$_3$OCHO          &  216.965900	& 5.5$\pm$0.4  & 0.8$\pm$0.1 & --- & ---  & --- & --- \\
CH$_3$OCHO          &  216.967994	& ---  & --- & 1.0$\pm$0.2 & 0.4$\pm$0.1  & --- & 1.1$\pm$0.2 \\
SiO         	    &  217.104984	& 7.0$\pm$0.2  &13.1$\pm$0.4 & 5.9$\pm$0.2 & 8.8$\pm$0.4  & 4.1$\pm$0.3 &14.2$\pm$0.5 \\
C$_2$H$_5$C$^{15}$N &  217.190187	& 2.4$\pm$0.2  & --- & --- & ---  & --- & --- \\
CH$_3$OCHO      	&  217.298353	& 3.2$\pm$0.2  & --- & --- & ---  & --- & 0.7$\pm$0.1 \\
CH$_3$OCHO          &  217.301326	& ---  & --- & 1.5$\pm$0.2 & 0.3$\pm$0.1  & --- & --- \\
$^{13}$CH$_3$OH     &  217.399550	& ---  & --- & --- & ---  & --- & --- \\
CH$_3$OH       	    &  217.418711	& ---  & --- & --- & ---  & --- & --- \\
($c-$C$_3$H$_2$)CH$_2$&  217.429650	& ---  & --- & --- & ---  & --- & 0.4$\pm$0.1 \\
\midrule 
 \multirow{2}*{Molecule} &  Frequency  & G351.77-0.54 & G353.41-0.36 & G035.13-0.74 & G034.41+0.24 & G034.26+0.15 \\                                                                                                        
	&    (GHz)    & [K km s$^{-1}$] &  [K km s$^{-1}$] & [K km s$^{-1}$] & [K km s$^{-1}$] &   [K km s$^{-1}$]     \\  
\midrule
CH$_3$OCHO          &  216.830197	& ---	& ---	& --- & --- & ---    \\
CH$_3$OCHO       	&  216.838889	& ---	& ---	& --- & --- & ---    \\
CH$_3$OH            &  216.945559	& 5.2$\pm$0.3	& ---	& --- & 2.9$\pm$0.1 & 0.5$\pm$ 0.1     \\
CH$_3$OCHO          &  216.965900	& ---	& ---	& --- & 1.5$\pm$0.1 & ---    \\
CH$_3$OCHO          &  216.967994	& 2.7$\pm$0.3	& ---	& --- & --- & ---    \\
SiO         	    &  217.104984	&24.0$\pm$1.3	& 3.8$\pm$0.2	& 1.4$\pm$0.1 & 5.4$\pm$0.2 & 2.6$\pm$ 0.2     \\
C$_2$H$_5$C$^{15}$N &  217.190187	& ---	& ---	& --- & --- & ---    \\
CH$_3$OCHO      	&  217.298353	& ---	& ---	& --- & 1.3$\pm$0.1 & ---    \\
CH$_3$OCHO          &  217.301326	& 2.4$\pm$0.2	& ---	& --- & --- & ---    \\
$^{13}$CH$_3$OH     &  217.399550	& 1.7$\pm$0.2	& ---	& --- & --- & ---    \\
CH$_3$OH       	    &  217.418711	& 2.2$\pm$0.3	& ---	& --- & --- & ---    \\
($c-$C$_3$H$_2$)CH$_2$&  217.429650	& ---	& ---	& --- & --- & ---    \\
\bottomrule                                                  
\end{tabular}\label{Tmb_Int}
\end{table*}  



\section{Analysis Method}


To propose molecular transitions candidates, we have taken the currently accepted procedure for line identification described by \citet{Herbst2009} to unequivocally identify new molecules: 
\textit{(i)} rest frequencies are accurately known.- To have accurately known frequencies, we only used molecular transitions in consolidated astronomical catalogs employing the spectroscopic database \textit{Splatalogue}\footnote{See: http://www.cv.nrao.edu/php/splat/}, which is a web-based platform that compiles information of several catalogs as The Cologne Database for Molecular Spectroscopy (CDMS, \citet{Muller2005JMoSt.742..215M}), Jet Propulsion Laboratory (JPL , \citet{Pearson2005IAUS..231P.270P}) and the Catalogue of National Institute of Standards and Technology (NIST, \citet{Lovas127721}), among others. 
\textit{(ii)} Observed frequencies of non-blended lines agree with rest frequencies for a single well-determined velocity of the source.- To do this,  we took the non-blended lines over the 3$\sigma$ level and measured the central frequency of their peak intensity, and tested if their velocity coincides with the $v_{LSR}$ of the source.
\textit{(iii)} All predicted lines of a molecule based on an LTE spectrum at a well-defined rotational temperature and appropriately corrected for beam dilution are present in the observed spectrum at roughly their predicted relative intensities.- This requirement was satisfied computing an XCLASS model \citep{xclass}. In this approach, we tested the proposed iCOMs candidates that fit the frequencies and simulate their emission spectra to compare with the observed spectra. Among the proposed iCOMS, the extra species were discarded by their discordance with the observed spectra, e.g., when a molecule generates \textit{artificial} lines that were not observed in experimental data (i.e. CH$_2$DOH,  (CH$_2$OH)$_2$, C$_3$H$_7$CN,  CHOCOOH or HOCHCHCHO). The final chemical diversity was tested by a classical non-parametric $\chi^{2}$ test whose results are presented in section \ref{sec:xclass}. 
\textit{(iv)} Simulate the entire spectrum of a source using all identified molecules to check consistency and separate contaminating blended features.- This step was also fulfilled with our XCLASS LTE synthetic spectrum, given that after the refinement of the species catalog we recompute the entire synthetic spectrum to compare it to the observed one.  Throughout this specific tunning, we also considered the blended species in the observed spectra, taking in consideration the proportions of the dominant and secondary iCOMs emitters according to their relative abundances to set up the modeling parameters.

The datacube of each source was centered at these non-blended lines, and their adjacent emission velocity channels were analyzed for congruent spatial distribution, which could be tracing the outflow lobes. 
To constrain the spatial distribution of the possible outflow component, we also computed maps of the integrated intensity in the blue- and red-shifted emission of the line profiles wider than FWZP $> 8$ kms$^{-1}$. These maps were plotted as intensity iso-contours over a background image composed using the color code: red, green, and blue, for the 3.6, 4.5, and 8.0$\mu$m bands of the IRAC camera of the SPITZER Space Telescope \citep{Cyganowski2008}. This allowed us to see the spatial correlation between the  4.5 $\mu$m band, the \textit{green fuzzy} emission which gives to EGOs their name \citep{Cyganowski} and the iCOMs spatial distribution. These maps are described and shown in section \ref{sec:integrated_maps}.

\subsection{XCLASS LTE Spectral Modeling}\label{sec:xclass}
To test our line identification method, we modeled the observed averaged spectra within the central 10$''$ of each source utilizing the XCLASS software. The computed grid was run with the following values as free parameters: T$_{rot}$ (rotational temperature in K), $N_{tot}$ (total column density in cm$^{-2}$), $v_{width}$ (velocity width in km s$^{-1}$) and $v_{off}$ (velocity offset in kms$^{-1}$). 

The T$_{rot}$ was limited from 10 to 500 K, the $v_{width}$ from 1 to 7 kms$^{-1}$ for the narrow emission lines and from 5 up to 15 kms$^{-1}$ for the wider ones. 
The $v_{off}$ range was from -2 to 2 kms$^{-1}$ away from the $v_{lsr}$ and the $N_{tot}$ values vary depending on the species, with the lowest and highest limits going from \num{1E13} to \num{1E19} cm$^{-2}$, in agreement with the typical values reported in other works \citep{Csengeri2019A&A, DeSimone2020A&A...640A..75D}. The source size parameter was fixed to 25$''$ as a conservative spatial extension covered by the iCOMs in our maps.  
The modeling of other simpler molecules was approached similarly, and their details will be discussed in a separate paper.

To establish the chemical diversity present in our bandwidth we took as archetype our source with more emission lines: G327.29-0.58. A sample plot of our resulting synthetic spectrum as well as the observed one is shown in Figure \ref{xclass_g327}. The complete collection of our XCLASS modeled spectra are displayed in Appendix \ref{sec:xclass_spec}. 

\begin{figure*}[htpb!]
	\centering  
	\includegraphics[width=\linewidth]{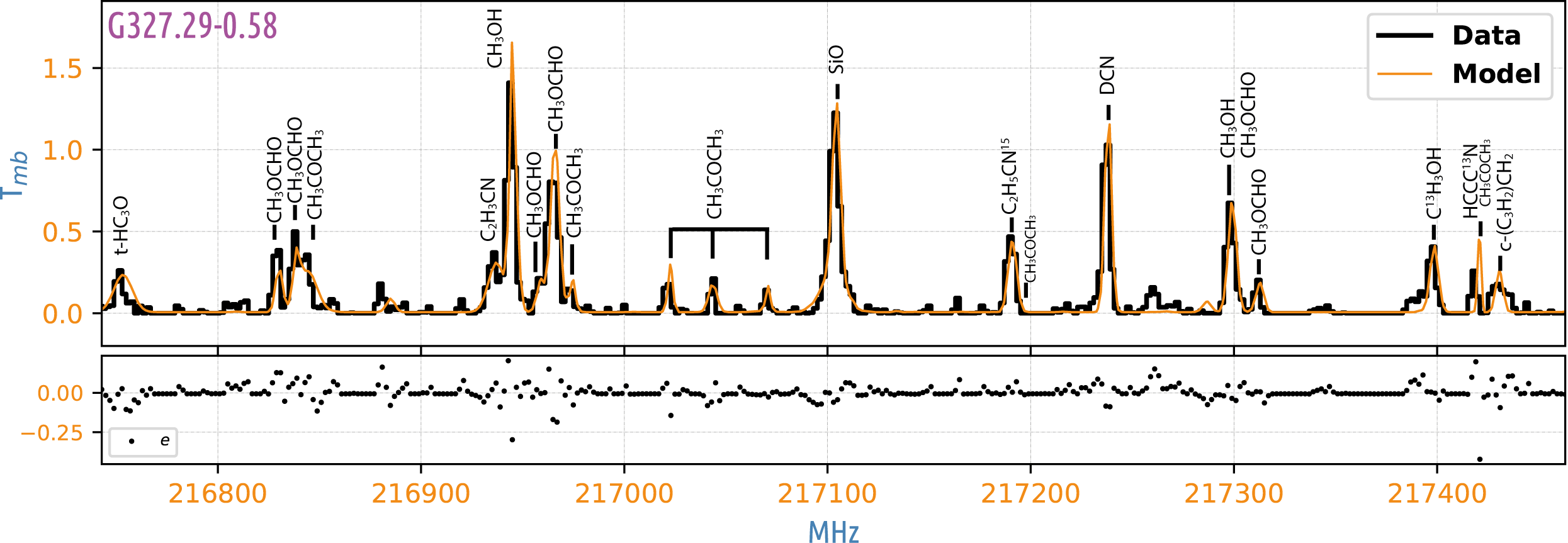}
	\caption{G327.29-0.58 observed spectra averaged inwards to the central $10''$ from our central position (black solid line) and their corresponding synthetic modeling using XCLASS software (orange solid line). 
	We also labeled the main contributing species to the spectra. In the lower subplot, we displayed the residuals ($e = I_{obs}-I_{model}$).}
	\label{xclass_g327}
\end{figure*}


The resulting physical parameters were computed using an algorithm chain, consisting of a particle swarm optimization method, named \textit{Bees} \citep{Bees2005}, linked to the \textit{Levenberg-Marquardt} method \citep{Marquardt1963}, an optimization of the Gauss-Newton and descendent gradient methods \citep{xclass}. Additionally, the error calculation was computed employing the \textit{Interval Nested Sampling} algorithm \citep{Ichida1979}. 

The goodness of the fit in our LTE models was evaluated by a classic $\chi ^{2}$ test. Using the respective channels in each bandwidth as our \textit{degrees of freedom}, we were able to compute a critical $\chi^{2}$  value to reach a p-value of 0.05 significance level in our entire sample. We assured that our modeled spectra successfully reached this critical $\chi^{2}$  value, therefore reinforcing the species identification.  


\section{Results: iCOMs emission towards massive protostars}
\subsection{XCLASS results}
Employing the LTE modeling, we were able to identify iCOMs emission in our sample within the frequency range 216.681 to 217.791 GHz. 
Our numerical approximation allowed us to compute $T_{rot}$ [K], $N_{tot}$ [cm$^{-2}$], and $v_{width}$ [kms$^{-1}$], for the detected iCOMs along our sample. We identified 3 complex molecules, from CH$_3$OH up to the more complex CH$_3$OCHO and ($c-$C$_3$H$_2$)CH$_2$, along with the two isotopologues $^{13}$CH$_3$OH and C$_2$H$_5$C$^{15}$N associated with the outflow stage in our sample. These findings are reported in Table \ref{xclass_result}.
\begin{table*}[!htpb]
	\centering
	
	\caption{Physical properties obtained with the XCLASS LTE modeling in our sample. For each detected species we reported the four fitted parameters (Top labeled): Rotational temperature [K], column densities  [cm$^{-2}$], Full Width at Half Power [km s$^{-1}$] and offset velocity [km s$^{-1}$]. If a source is not listed below, the iCOM namely was non-detected. The \textit{labeled *)} $\Delta v$ results shows the line widths smaller than our spectrometer resolution, and it were fitted in cases in which the emission are blended and therefore the lines contribute to conform a wider line.}
	\begin{tabular}{lcccclcccc}
		\toprule
		 \multirow{3}*{Source} & T$_{rot}$ & N$_{Tot}$  & $\Delta v$& $v_{off}$  & \multirow{3}*{Source} &  T$_{rot}$ & N$_{Tot}$  & $\Delta v$ & $v_{off}$\\
		                       & [K]       & [cm$^{-2}$] & [kms$^{-1}$] & [kms$^{-1}$]                   &       &  [K]       & [cm$^{-2}$] & [kms$^{-1}]$ & [kms$^{-1}$]\\
		\cmidrule(r){2-5}                                                 \cmidrule(r){7-10} 
		                       &\multicolumn{4}{c}{CH$_3$OH}      &     & \multicolumn{4}{c}{CH$_3$OCHO}      \\
		\cmidrule(r){1-5}                                                 \cmidrule(r){6-10} 
G03426$+$015 & 	177$\substack{- 78 \\+  92}$& 2.4$\substack{-0.9\\+3.7}\times10^{15}$&  3$\substack{-2 \\+ 2}$ & -1$\substack{-1 \\+ 1}$& G03426$+$015 & 	 40$\substack{- 18\\+21}$& 5.7$\substack{-3.7\\+7.4}\times10^{13}$& 1$\substack{-0\\+ 2}$* & -1$\substack{-1\\+ 1}$\\
G03441$+$024 & 	225$\substack{- 23 \\+  44}$& 5.1$\substack{-3.3\\+5.0}\times10^{15}$&  5$\substack{-3 \\+ 1}$ & -1$\substack{-1 \\+ 1}$& G03441$+$024 & 	 59$\substack{- 24\\+16}$& 2.7$\substack{-2.2\\+1.9}\times10^{14}$&   7$\substack{-4\\+ 0}$ & -1$\substack{-1\\+ 1}$\\
G03513$-$074 & 	255$\substack{-100 \\+  71}$& 2.7$\substack{-1.6\\+3.1}\times10^{15}$&  4$\substack{-2 \\+ 1}$ & -1$\substack{-1 \\+ 1}$& G03513$-$074 & 	 39$\substack{- 18\\+19}$& 7.6$\substack{-5.3\\+9.0}\times10^{13}$& 2$\substack{-1\\+ 1}$* &  1$\substack{-2\\+ 1}$\\
G32729$-$058 & 	289$\substack{- 26 \\+  24}$& 4.0$\substack{-3.6\\+7.7}\times10^{15}$&  5$\substack{-2 \\+ 1}$ &  0$\substack{-1 \\+ 1}$& G32729$-$058 & 	 75$\substack{- 13\\+54}$& 3.6$\substack{-2.1\\+1.1}\times10^{16}$&   5$\substack{-1\\+ 1}$ &  0$\substack{-1\\+ 1}$\\
G32881$+$063 & 	324$\substack{- 31 \\+  20}$& 4.1$\substack{-3.4\\+2.5}\times10^{16}$&  5$\substack{-2 \\+ 1}$ &  0$\substack{-1 \\+ 1}$& G32881$+$063 & 	400$\substack{-  0\\+35}$& 1.7$\substack{-1.6\\+2.0}\times10^{15}$& 2$\substack{-1\\+ 1}$* & -2$\substack{-0\\+ 1}$\\
G33313$-$043 & 	187$\substack{- 54 \\+  44}$& 1.9$\substack{-0.7\\+0.9}\times10^{16}$&  8$\substack{-3 \\+ 1}$ & -2$\substack{-0 \\+ 1}$& G33313$-$043 & 	100$\substack{-  0\\+93}$& 4.1$\substack{-3.4\\+3.0}\times10^{14}$&   3$\substack{-2\\+ 1}$ & -2$\substack{-0\\+ 1}$\\
G33792$-$048 & 	308$\substack{- 17 \\+  22}$& 1.1$\substack{-0.8\\+0.6}\times10^{16}$&  5$\substack{-3 \\+ 1}$ & -1$\substack{-1 \\+ 1}$& G33792$-$048 & 	403$\substack{-  3\\+42}$& 2.5$\substack{-0.9\\+1.2}\times10^{15}$&   6$\substack{-3\\+ 1}$ & -1$\substack{-1\\+ 1}$\\
G34097$-$102 & 	165$\substack{- 33 \\+  31}$& 4.5$\substack{-2.2\\+1.6}\times10^{15}$&  5$\substack{-3 \\+ 1}$ & -1$\substack{-1 \\+ 1}$& G35145$+$066 & 	111$\substack{- 11\\+25}$& 5.5$\substack{-4.6\\+3.5}\times10^{14}$&   3$\substack{-2\\+ 1}$ &  0$\substack{-1\\+ 1}$\\
G35145$+$066 & 	160$\substack{- 38 \\+  52}$& 1.9$\substack{-1.0\\+0.7}\times10^{16}$&2$\substack{-1 \\+ 2}$* & -1$\substack{-1 \\+ 1}$& G35177$-$054 & 	404$\substack{- 57\\+56}$& 7.9$\substack{-6.7\\+2.1}\times10^{15}$&   7$\substack{-4\\+ 0}$ & -2$\substack{-0\\+ 1}$\\
G35177$-$054 & 	126$\substack{- 16 \\+  34}$& 2.0$\substack{-1.2\\+0.7}\times10^{16}$&  7$\substack{-3 \\+ 1}$ & -1$\substack{-1 \\+ 1}$\\
G35341$-$036 & 	362$\substack{-191 \\+  54}$& 7.2$\substack{-5.2\\+6.1}\times10^{15}$&  4$\substack{-2 \\+ 1}$ &  0$\substack{-1 \\+ 1}$\\

		\cmidrule(r){1-5}                                                 \cmidrule(r){6-10} 
		Source     &\multicolumn{4}{c}{$^{13}$CH$_3$OH}       & Source     & \multicolumn{4}{c}{CH$_3$COCH$_3$}      \\
		\cmidrule(r){1-5}                                                                  \cmidrule(r){6-10} 
G32729$-$058 & 	 57$\substack{- 23 \\+  19}$& 1.6$\substack{-0.1\\+0.9}\times10^{16}$&  6$\substack{-2\\+ 1}$ &  1$\substack{-2\\+ 1}$& G03426$+$015 & 	 10$\substack{-  0\\+  97}$& 0.2$\substack{-0.0\\+1.1}\times10^{14}$& 1$\substack{-0\\+ 2}$* & -2$\substack{ 0\\+ 2}$\\
G33792$-$048 & 	 22$\substack{-  8 \\+   4}$& 2.5$\substack{-1.0\\+1.2}\times10^{17}$&  5$\substack{-6\\+ 0}$ & -2$\substack{-0\\+ 1}$& G32729$-$058 & 	153$\substack{- 26\\+  20}$& 1.7$\substack{-0.2\\+1.4}\times10^{15}$&   3$\substack{-0\\+ 1}$ &  0$\substack{-1\\+ 1}$\\
G35177$-$054 & 	 20$\substack{-  7 \\+   0}$& 1.0$\substack{-1.0\\+0.0}\times10^{08}$& 10$\substack{-6\\+ 0}$ &  2$\substack{-2\\+ 0}$\\
G35341$-$036 & 	 56$\substack{- 46 \\+ 126}$& 5.7$\substack{-4.2\\+7.2}\times10^{15}$& 10$\substack{-6\\+ 0}$ & -2$\substack{-0\\+ 1}$\\                                                       

		\cmidrule(r){1-5}                                                                  \cmidrule(r){6-10} 
		Source   &\multicolumn{4}{c}{C$_2$H$_3$CN}                 & Source     & \multicolumn{4}{c}{C$_2$H$_5$C$^{15}$N}      \\
		\cmidrule(r){1-5}                                                                  \cmidrule(r){6-10} 
G32729$-$058 & 	474$\substack{- 27 \\+  14}$& 1.6$\substack{-0.1\\+1.2}\times10^{15}$& 3$\substack{-4\\+ 1}$ & -1$\substack{-1 \\+ 1}$& G03441$+$024 & 	 14$\substack{-  4 \\+  19}$& 8.2$\substack{-7.1\\+1.8}\times10^{14}$& 7$\substack{-4\\+ 1}$ & -2$\substack{ 0 \\+ 1}$\\
G35177$-$054 & 	301$\substack{-  1 \\+  86}$& 3.7$\substack{-2.2\\+2.4}\times10^{14}$& 3$\substack{-2\\+ 2}$ & -1$\substack{-2 \\+ 1}$& G32729$-$058 & 	 16$\substack{-  5 \\+   0}$& 6.4$\substack{-5.4\\+3.6}\times10^{15}$& 6$\substack{-2\\+ 1}$ & -1$\substack{-1 \\+ 1}$\\
		\cmidrule(r){1-5}                                                                  \cmidrule(r){6-10} 
		Source     & \multicolumn{4}{c}{($c-$C$_3$H$_2$)CH$_2$;v=0}  & & & &    \\
		\cmidrule(r){1-5}
G32729$-$058 & 	324$\substack{- 24 \\+  50}$& 2.4$\substack{-2.0\\+1.8}\times10^{17}$& 5$\substack{-3 \\+ 0}$ & -1$\substack{-1 \\+ 1}$\\
\bottomrule    
	\end{tabular}\label{xclass_result}
\end{table*}



\subsection{Integrated Maps results}\label{sec:integrated_maps}
\begin{figure}[htpb!]
	\centering  
	\includegraphics[width=\linewidth]{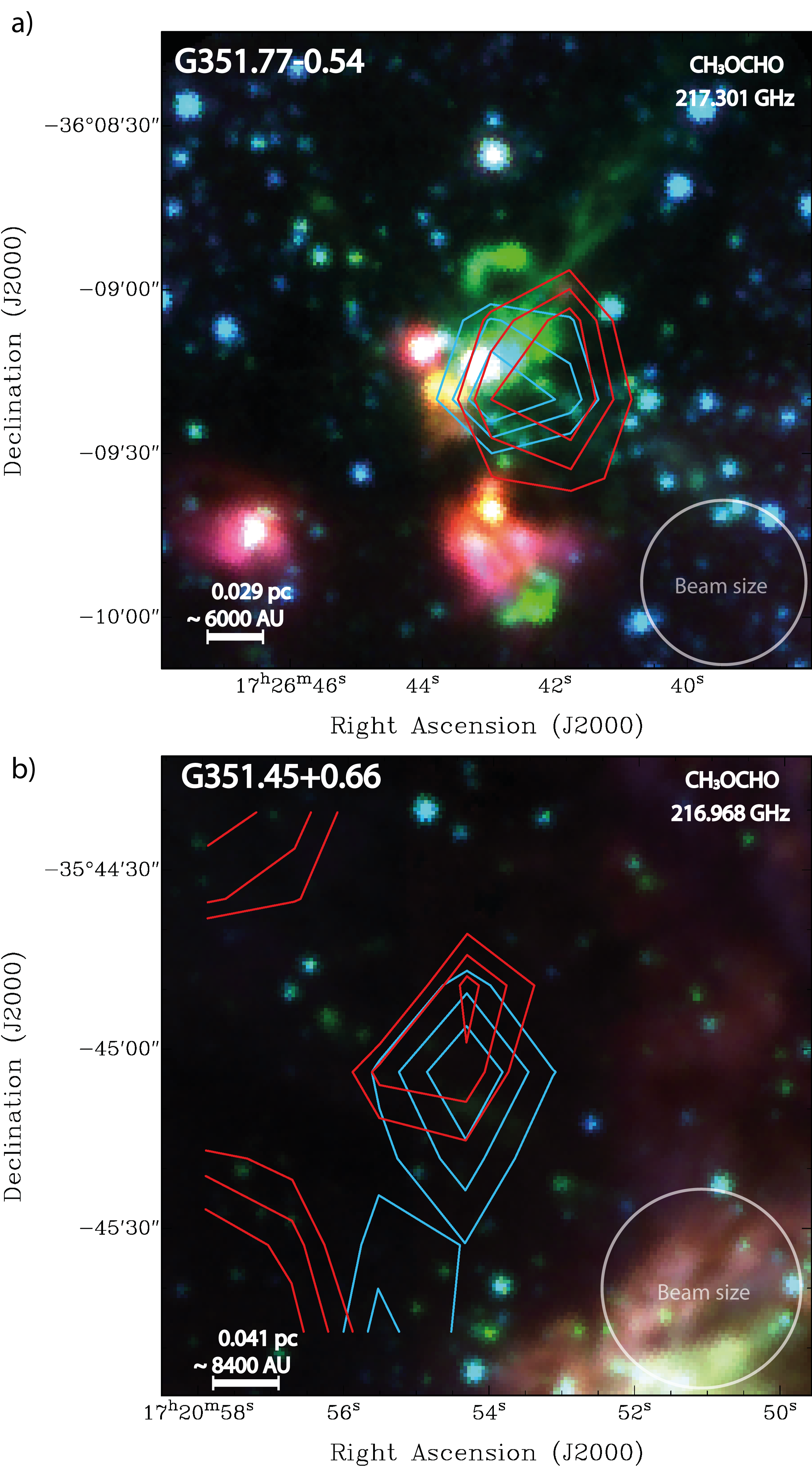}
	\caption{Contour maps of the blue- and red-shifted emission of MF at \textit{a)} G351.77-0.54 and \textit{b)} G351.45+0.66. The respective central frequency and physical scale are labeled. The contour maps show the emission up from 66\%+10\% peak emission, the blue and red contours stands for the blue- and red-shifted emissions, respectively. The shifted emissions were integrated outward from the central 8 km s$^{-1}$ to reach the FWZP for each emission line.}
	\label{MF_Figure_1}
\end{figure}

In order to elucidate the nature of the emission, we inspected the integrated maps of the shifted emission in the broadest iCOMs, looking for high-velocity spatial components. For this, we computed the integrated maps of the blue- and red-shifted emissions for iCOMs with a FWZP wider than 8 kms$^{-1}$ ($|v\footnotesize{\text{(min)}}_{wing}-v_\text{LSR}| \geq 4.0$ kms$^{-1}$). This criterion, corresponding to FWHM$\sim3$\,kms$^{-1}$ (for a Gaussian profile assuming FWZP$\sim6\sigma$), is appropriate as it would include previously observed iCOMs associated with outflow environments (with FWZP$\gtrsim 10$\, km\,s$^{-1}$, or FWHM$\gtrsim 3$\, km\,s$^{-1}$, e.g. \citet{Palau2007A&A...465..219P,Codella2015MNRAS.449L..11C,Burkhardt2016ApJ...827...21B}). This procedure tries to avoid excitation mechanisms related to the central core \citep{Walmsley1999,Schilke2001A&A...372..291S} and reinforce that the emission is produced as a result of outflow activity \citep{tafa2013active,Li2019ApJ...878...29L}. Thus, looking for elongated distributions along adjacent channels in the datacube could diagnose the emission related to large scale outflows. This elongation should be more extended than the $\theta_{\text{HPBW}}$ for each telescope ($\sim 30.5''$ for APEX and $\sim 11.94''$ for IRAM) and the minimum detectable offset should be greater than three times the pointing accuracy ($12''$ for APEX and $6''$ for IRAM).
After this analysis, we have not found collimated outflow structures traced by the iCOMs. However, we point out that this could be a resolution problem, as the angular resolution reached in our mapping observations is still poor and beam dilution could be affecting the compact and faint structures expected from outflows \citep{Csengeri2019A&A}. Nevertheless, we have found several iCOMs maps with the blue and red-shifted emission peaking at offset positions (away from the near-IR peak). This spread emission could be tracing large-scale low-velocity movements, possibly related to bipolar or multipolar outflows. The most prominent example of this behavior is the methyl formate (MF, here and after) emission at 217.301 GHz towards \textit{G351.77-0.54}, which shows an east-west elongated emission, a spatial distribution also observed in its methanol emission (see Figure \ref{MF_Figure_1}, a). Nontheless,  we should recall that iCOM emission does not strictly follow the EGO emission, but this association could be plausible considering our low resolution and the spatially spread emission at 4.5 $\mu$m shown by this object. 

The MF emission at 216.968 is also interesting towards \textit{G351.45+0.66}, since the blue- and red-shifted contour maps show an offset of their peaks, following a north-south direction (see Figure \ref{MF_Figure_1}, b). This behavior is also followed by the methanol emission at 216.945 GHz and the MF at 217.298 GHz.

\begin{figure}[htpb!]
	\centering  
	\includegraphics[width=\linewidth]{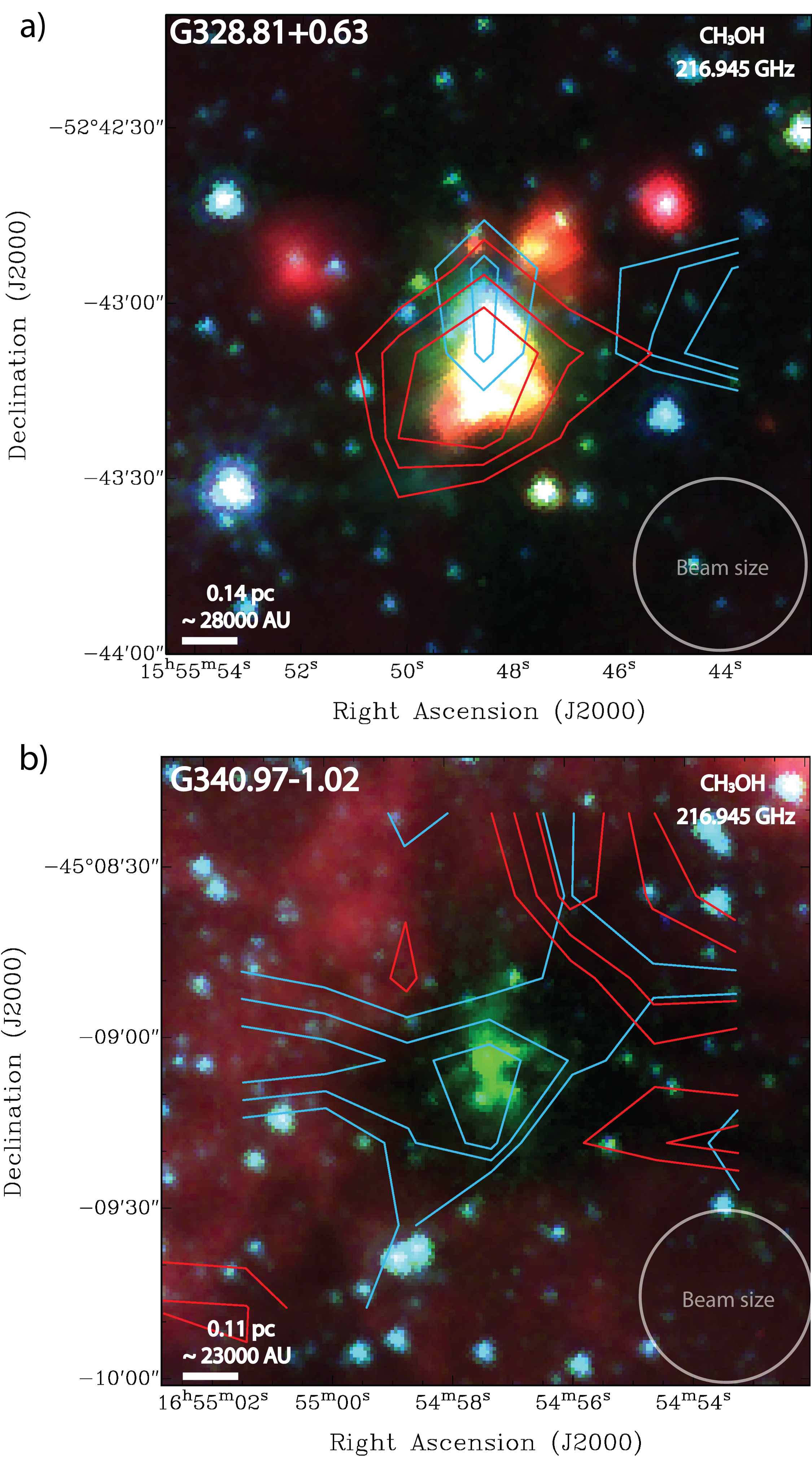}
	\caption{Contour maps of the blue- and red-shifted emission of methanol at \textit{a)} G328.81+0.63 and \textit{b)} G340.97-1.02. The respective central frequency and physical scale are labeled. The contour maps show the emission up from 66\%+10\% peak emission, the blue and red contours stands for the blue- and red-shifted emissions, respectively. The shifted emissions were integrated outward from the central 8 km s$^{-1}$ to reach the FWZP for each emission line.}
	\label{Meth_Figure_2}
\end{figure}

The object \textit{G328.81+0.63} also shows an offset of the blue- and red-shifted methanol emission at 216.945 GHz, in the north-south direction (Figure \ref{Meth_Figure_2}, a).

The blue-shifted emission traced by methanol at 216.945 GHz towards \textit{G340.97-1.02} shows a peak following the southeast-northwest direction of the EGO, but at a larger scale (Figure \ref{Meth_Figure_2}, b). These morphological features are potential candidates for high angular resolution interferometric studies, to avoid the beam dilution and therefore tightly constrain the extended emission of this iCOM. 

\begin{figure*}[htpb!]
	\centering  
	\includegraphics[width=0.85\linewidth]{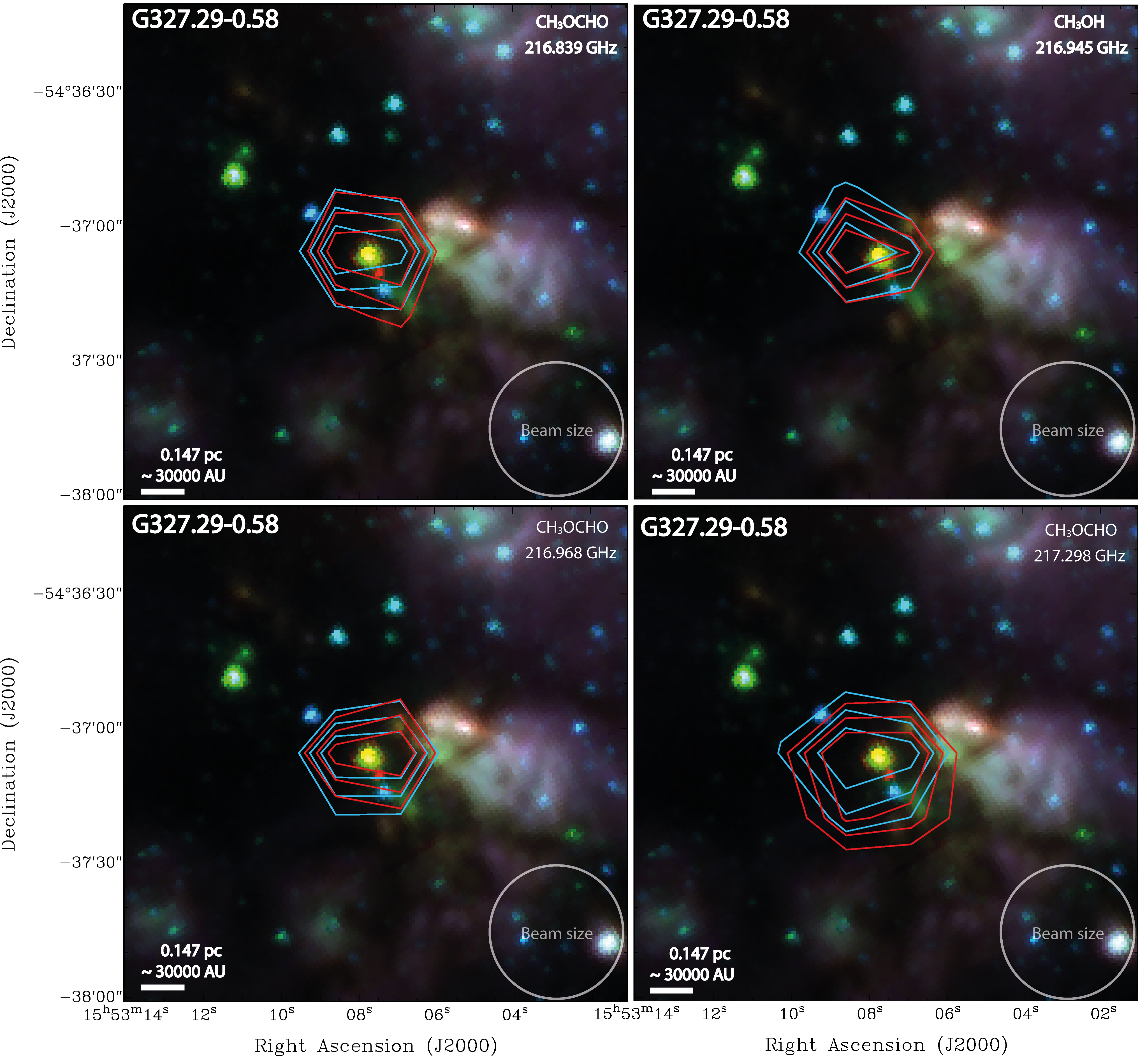}
	\caption{Contour maps of the blue- and red-shifted emission for several iCOMs detected in G327.29-0.58. The corresponding specie and its central frequency is labeled in each frame. The contour maps show the emission up from 66\%+10\% peak emission, the blue and red contours stands for the blue- and red-shifted emissions, respectively. The shifted emissions were integrated outward from the central 8 km s$^{-1}$ to reach the FWZP for each emission line.}
	\label{Figure_3}
\end{figure*}

Other iCOM emission lines are mainly tracing the massive protostar but their closest intensity contour peaks, for both blue- and red-shifted emission, are commonly overlapped on the inner parts of the core, and therefore related to the extended envelope. This behaviour is appreciated in G327.29-0.58, as their iCOMs contour maps shows an compact emission that are related to the core and extends up to outer envelope \ref{Figure_3}. Even when the contours shows a faint east-to-west  elongation this appreciation is marginal at our angular resolution. 

The whole set of previously mentioned integrated maps are displayed  in Figure \ref{Maps}. 

\subsection{iCOMs correlations}
\begin{figure}[htpb!]
	\centering  
	\includegraphics[width=\linewidth]{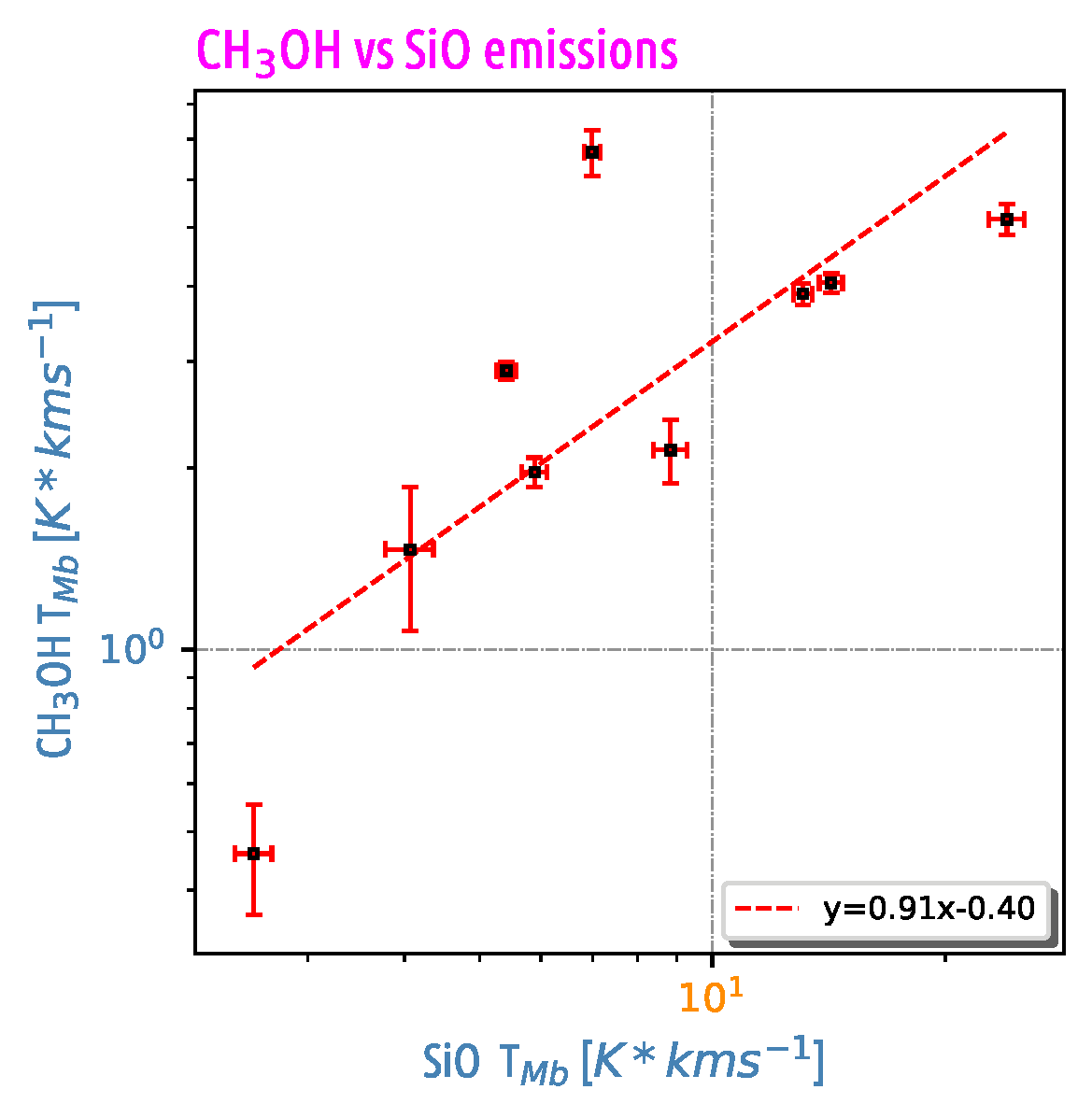}
	\caption{Dispersion of the integrated emission of CH$_3$OH at 216.945559 GHz versus the shock tracer SiO. The lower-right label shows the values for a linear fit in the log-log space. The upper outlier is the most chemically active source in our sample G327.29-0.58. These correlation points out that the shock environment in massive protostars favours the CH$_3$OH gas-phase population.} 
	\label{ch3ohSiO}
\end{figure}

The most prominent and non-blended observed iCOM in our bandwidth was the methanol transition at 216.945 GHz, which was observed in all but G035.13-0.74 and G353.41-0.36. From this, we have proposed to test the correlation between their integrated emission and the \textit{shock tracer} SiO at 217.105 GHz (also present within our bandwidth). We tested their correlation by employing two non-parametric tests, the Pearson and Spearman's Rank correlations, in order to evaluate a linear or a monotonic dependence, respectively. The analysis showed a Spearman Rank correlation factor $r_S$ of 0.76, with a corresponding p-value of 0.016, and a Pearson coefficient $r_P$ of 0.58, with a p-value of 0.10. These two tests have confirmed the relation between these emission lines with a good significance level. And, given that these two lines were observed simultaneously, the systematic errors are expected to affect similarly to both emission lines, reinforcing this correlation.  We plotted this distribution and used a linear regression in log-log space to illustrate this trend, obtaining a slope close to unity i.e. $y=0.91x-0.40$, as it is displayed in Figure \ref{ch3ohSiO}. The only outlier is G327.29-0.58, an ATLASGAL object that differs from the other regions of the sample as it is not an extended object at 4.5 $\mu$m, but rather a compact near-IR bright object. This together with a higher chemical activity suggest a late evolutionary stage for G327.29-0.58.

The observed correspondence suggests the enhancement of CH$_3$OH within the shocks. This relation could also be explained because shock conditions boost the desorption of iCOMs, previously synthesized onto the dust and ice mantles.

The efficient release of the CH$_3$OH parent molecule to the gas phase could ease other gas-phase reactions and therefore lead to secondary molecules as methyl formate and dymethil ether \citep{Balucani2015}, iCOMs that are feasible to be formed at the temperatures we have measured towards our sample, $\sim$ 30 K \citep{Minissale2016A&A...585A..24M, Ceccarelli2017ApJ...850..176C}. The injection of kinetic energy into the grains not only boosts the evaporation through thermal desorption, but also the mobility of radicals in the grains, stimulating movements of the radicals in the dust through the Langmuir-Hinshelwood mechanism \citep{Herbst2009}. 

Due to the essential role of the evolution of methanol as a precursor of several iCOMs, we analyzed the CH$_3$OH/SiO ratio against the popular time indicator ratio $L_{Bol}/M_{dust}$, expecting a decreasing trend as the source evolves \citep{Lopezsepulcre2012,Sanchezmongue2013,Csengeri2016}. 
In this case, we take the values of $L_{bol}$ and $M_{dust}$ from literature \citep{miettinen2006, urquhart2014, wienen2015, Csengeri2016}; but, as L$_{bol}$ was not reported for all our sources, this results in a sample size of six sources, therefore, we were unable to establish any conclusive result. Nevertheless, we think this relation is a way to measure the efficiency of the shocks releasing the CH$_3$OH into the gas phase as time evolves.



Similarly, we also evaluated the CH$_3$OH/SiO ratio versus the dust mass of the object using the same tests. In this case, we found dust mass data for nine of our sources, which traces an increasing trend with a corresponding correlation values of $r_S=0.19$; p-value $ =0.61$, and $r_P=0.19$; p-value $=0.62$. 
Both of these correlation coefficients are low, but due to our low significance p-values, we can not fully establish this null relation.



\section{Discussion}\label{sec:discussion}


\subsection{Methanol}
The methanol molecule was found in 100$\%$ of our sample. This parent molecule in its gas phase is the precursor of several iCOMs \citep{Ceccarelli2017ApJ...850..176C}. It has been proved to be formed on the grain surfaces by hydrogenation of frozen CO by the successive addition of H atoms  \citep{Boogert2015ARA&A..53..541B, Tielens1982A&A...114..245T, Watanabe2002ApJ...571L.173W}. 
The main proposed mechanisms to release this iCOM into the gas phase are cosmic rays or thermal desorption \citep{Ceccarelli2017ApJ...850..176C}.  In our sample, we expect that the shocks themselves could transfer the needed amount of energy into the dust grains to induce evaporation. 

Our observed methanol emission lines suggest a high temperature in the gas induced by the shocks, with values going from $126$ to $362\ K$ and an average value of $234\pm78\ K$. This temperature is close to the dust temperature reported by \citet[][]{Fu2016RAA....16..182F} of 250 K towards G34.26+0.15. 

These high temperatures along with the observed iCOM chemical richness in our sample are congruent with a C-type shock \citep{arce,palau2017}.

To test the shock impact into the CH$_3$OH enhancement, we examined the relation between the T$_{mb}$ integrated emission of CH$_3$OH at 216.945 GHz and SiO at 217.105 GHz. In our sample, this shows an increasong behavior described by a linear Pearson correlation with r$_P$ of 0.78 and a p-value of 0.013. This relation suggests the efficiency of shocks boosting methanol to their gas phase. Considering the importance of grain-surface chemistry in the methanol formation \citep{Garrod2006A&A...457..927G,Balucani2015} we could speculate that the observed methanol enhancement in our sample implies an increase in the dust mass (a property of ATLASGAL sources), which allows the shocks to easier injects the energy needed to sublimate the methanol from the dust grains. 
This reinforces the hypothesis that the radiation process is not the sole necessary mechanism to generate complex molecules near to forming stars \citep{Arce2008,Zeng2018MNRAS.478.2962Z}. 

Alternatively, this could suggest that SiO and CH$_3$OH are being affected by a common excitation mechanism as photo-evaporation due to an external UV field \citep{Schilke2001A&A...372..291S,Palau2007A&A...465..219P}. In that case, we should also consider and further discriminate if the central integrated line of SiO is under the influence of another excitation mechanism different to shocks, but this topic is beyond the scope of this study.

Methanol abundance enhancement has been observed in several outflows \citep{Bachiller95,Palau2007A&A...465..219P}, and their line profile wings are significantly lower than that of SiO. This could be due to CH$_3$OH not surviving at velocities as high as those reached by the outflow, or to the fact that high-velocity shocks are required to boost the injection of Si into the gas phase \citep{Guillet2011A&A...527A.123G}.

According to our maps, the spatial distribution of methanol in our sample follows the central 4.5 $\mu$m peak emission, and for most of the cases sustains a nearly circular morphology. Their extension is wide and seems to be related to the central massive protostar, tracing the central $\sim 20''$ away from the central pointing and up to $\sim 35''$ for the more extended sources. This represents a physical scale going from $\sim$ 0.1 to 0.5 pc according to the distance of the objects.

In G328.81+0.63, the methanol red-shifted emission displays an elongated southeast to northwest morphology, which covers $\sim 35''$ ($\sim$98$\times10^{3}$ AU) (Figure \ref{Meth_Figure_2}, a). On the other hand, the blue-shifted emission is located along the north-south axis and has a more compact structure. The emission center of these two components shows an offset, which suggests a relationship with an extended component. Further observations with improved angular resolution are necessary to confirm the outflow origin of this molecule.

In G340.97-1.02, the methanol blueshifted emission at 216.945 GHz is elongated in the southeast-northwest direction for more than $\sim 30''$ (Figure \ref{Meth_Figure_2}, b). 

\begin{figure}[htpb!]
	\centering  
	\includegraphics[width=\linewidth]{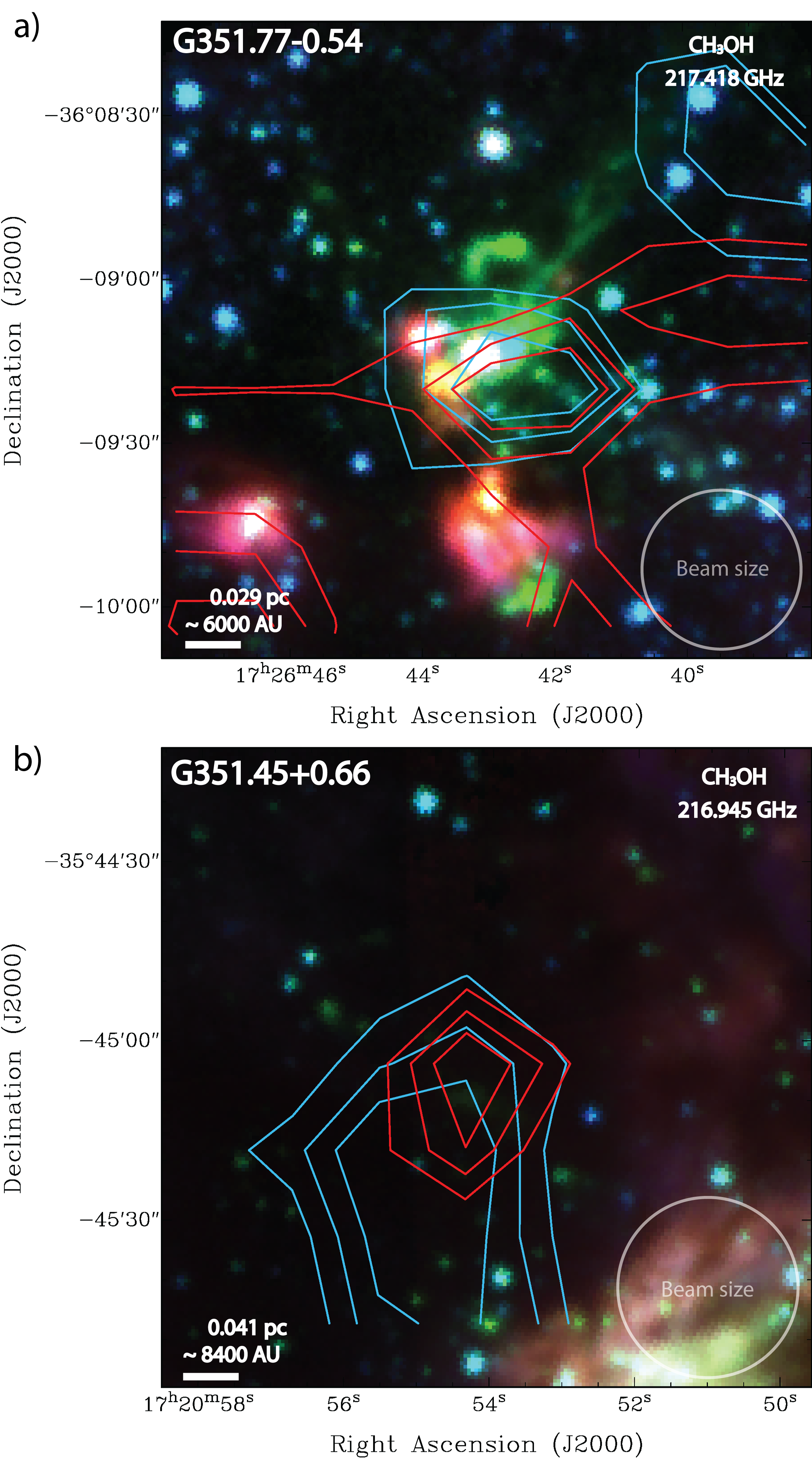}
	\caption{Contour maps of the blue- and red-shifted emission of methanol at a) G351.77-0.54 and b) G351.45+0.66. Central frequency and physical scale are labeled. The contours show the emission up from 66\%+10\% peak emission, the blue and red contours stands for the blue- and red-shifted emissions, respectively. The shifted emissions were integrated outward from the central 8 km s$^{-1}$ to reach the FWZP in the emission line.}
	\label{methanol_wings}
\end{figure}

Towards G351.77-0.54, the methanol red-shifted emission at 217.418 GHz, spreads along the east-west direction for $\sim 80''$ ($\sim$24$\times10^{3}$ AU), following the central EGO emission (Figure \ref{methanol_wings}, a). Also, the red-shifted emission extends to the south, but this emission is probably related with the compact nearest 4.5 $\mu$m source located just 35$''$ to the south. 

In G351.45+0.66, the methanol blue-shifted emission at 216.945 GHz is directed to the southern part from the central pointing for more than $\sim 40''$  (Figure \ref{methanol_wings}, b). This emission is poorly collimated  and it could be more related to the dust continuum emission.

\subsection{Methyl Formate}
Methyl formate was found in $\sim$ 82$\%$ of our sample. This molecule is expected to be formed on ices by the reaction of CH$_3$O with HCO and O \citep{Ceccarelli2017ApJ...850..176C}, or in the gas phase reactions starting from dimethyl ether \citep{Balucani2015}. Even at our high detection rate, the correlation of this molecule versus the previously mentioned tracer was avoided as their emission lines among the sources belong to different transitions, and this implies a non-homogeneous correlation bias. 

The lower limit of the computed range of rotational temperatures ($40-404$ K) is close to the temperatures required by gas-grain chemical models as in \citet[][$T_{rot} \sim$ 30 K]{Garrod2006A&A...457..927G}, which explains the formation of iCOMs such as CH$_3$OCH$_3$, CH$_3$OCHO, and CH$_3$CHO. This temperature is needed for the radicals to have sufficient energy to ensure their mobility \citep{Bacmann2012A}.  Even when this model does not explain the presence of these iCOMs in their gas phase, as previously mentioned in our sample the shocks could be a major contributor to this release. 


It is noteworthy that the XCLASS fit gives an average MF $T_{rot}$ value of 184 K, i.e. more than the temperature needed to trigger the gas-phase reactions \citep{Minissale2016A&A...585A..24M,Ceccarelli2017ApJ...850..176C}. 

In $\sim 45$\% of the sample, the MF emission presents an FWZP $\sim $ 6 kms$^{-1}$, not broader enough to overpass our limit to define an outflow origin of the emission. Actually, the integrated emission shows a circular morphology centered on the central massive protostar with an average coverage of 15$''$.

Contrarily, the following examples have shown an outstanding spatial distribution in the MF emission.  


In G351.77-0.54 the red-shifted emission is extended and seems to follow the eastern emission of the EGO (Figure \ref{MF_Figure_1}, a); however, the presence of nearby sources could affect the emission and produce the observed broadening. 

The MF emission in G351.45+0.66 at 216.968 GHz shows an offset location of their blue- and red-shifted peaks. These peaks are located in a north-south direction through a $\sim 35''$ extension (Figure \ref{MF_Figure_1}, b), and, in contrast to the other sources in our sample, they do not have strong 4.5 $\mu$m emission, being possibly the less evolved source in our sample (note, however, that this source was selected from the ATLASGAL catalog). The blue-shifted emission traced by MF is more compact and elongated than the blue-shifted emission of methanol (Figure \ref{methanol_wings}, b). 

\begin{figure}[htpb!]
	\centering  
	\includegraphics[width=\linewidth]{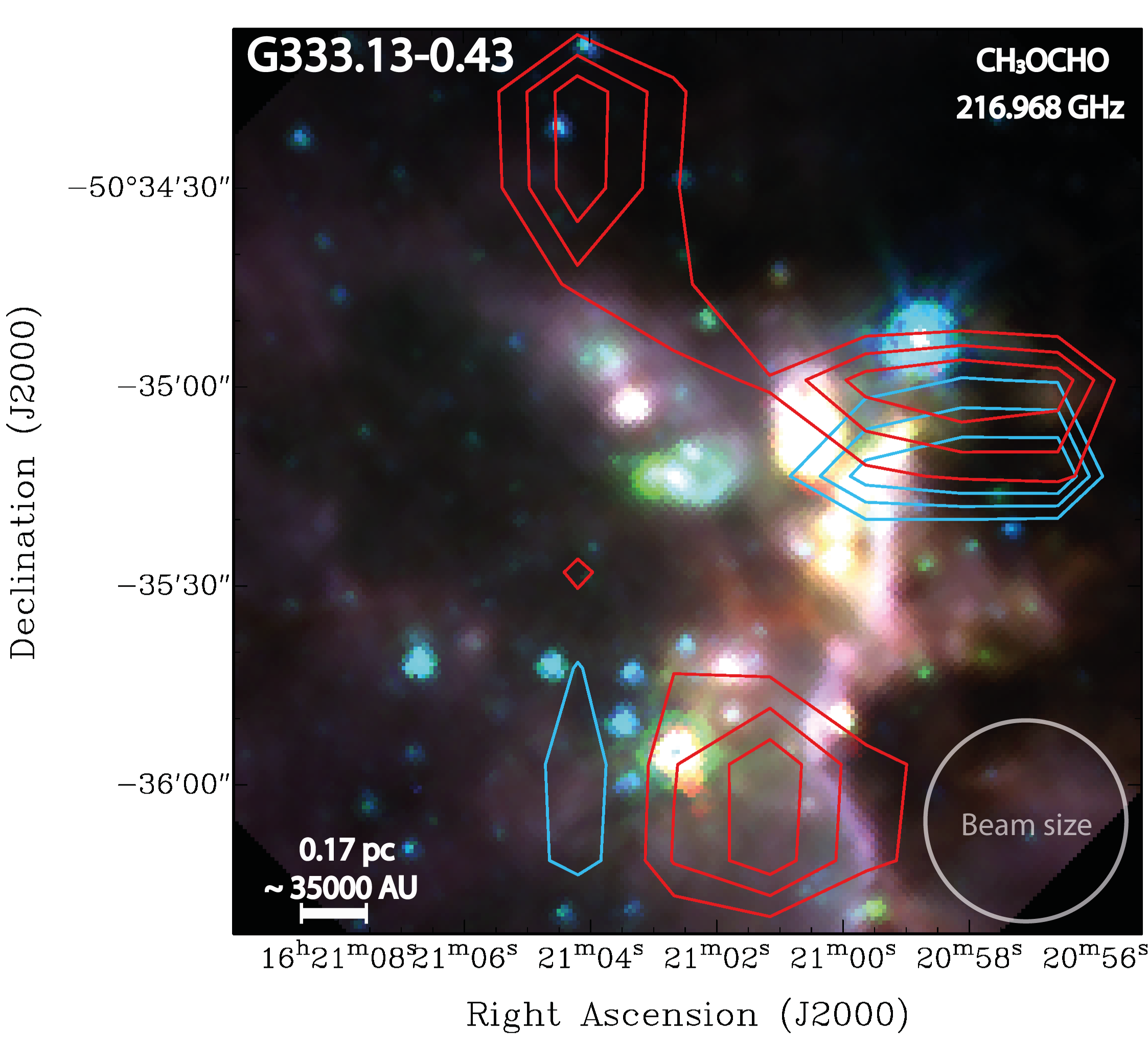}
	\caption{Contour maps of the blue- and red-shifted emission of MF at G333.13-0.43. Central frequency and physical scale are labeled. The contours show the emission up from 66\%+10\% peak emission, the blue and red contours stands for the blue- and red-shifted emissions, respectively. The shifted emissions were integrated outward from the central 8 km s$^{-1}$ to reach the FWZP for each emission line.}
	\label{MF_wings}
\end{figure}

In G333.13-0.43, the blue- and red-shifted emission is completely offset from our central pointing (see Figure \ref{MF_wings}).  The western emission has an elongated bipolar morphology, which could be tracing an outflow. Due to the fact that this emission is located $\sim15''$ away from the central object and the fact that the near-IR emission is very complex, this MF emission could be associated with nearby sources different to the massive protostar studied here. 

\subsection{Acetone}
Acetone was detected in G034.26+015 and G327.29-0.58. Their emission lines were confirmed by the XCLASS model with three unblended transitions at 216.972, 217.021, and 217.070 GHz, and a faint contribution at 217.191 GHz. In all the cases the line intensity is faint and is tracing the central source. The contour maps shows a circular extension over $\sim35''$ 

Its presence in our XCLASS modeling is justified as it contributes to the optimal fit for faint emission lines in the chemical rich object G327.29-0.58, improving the significance of the $\chi^2$ test.

\subsection{Ethyl cyanide}
The ethyl cyanide (C$_2$H$_5$C$^{15}$N) isotopologue has been detected in G034.41+0.24 and G327.29-0.58, with two transitions within our bandwidth: one at 216.880 GHz ($S/N\sim 3\sigma$ in G327.29-0.58, not present in G034.41+0.24), and another at 217.190 GHz ($S/N\sim4\sigma$ in G034.41+0.24 and $S/N\sim6\sigma$ in G327.29-0.58). Note, however, that the last emission line is blended with a faint CH$_3$COCH$_3$ signature, but this is just a minor contributor to the total observed intensity.

\begin{figure}[htpb!]
	\centering  
	\includegraphics[width=\linewidth]{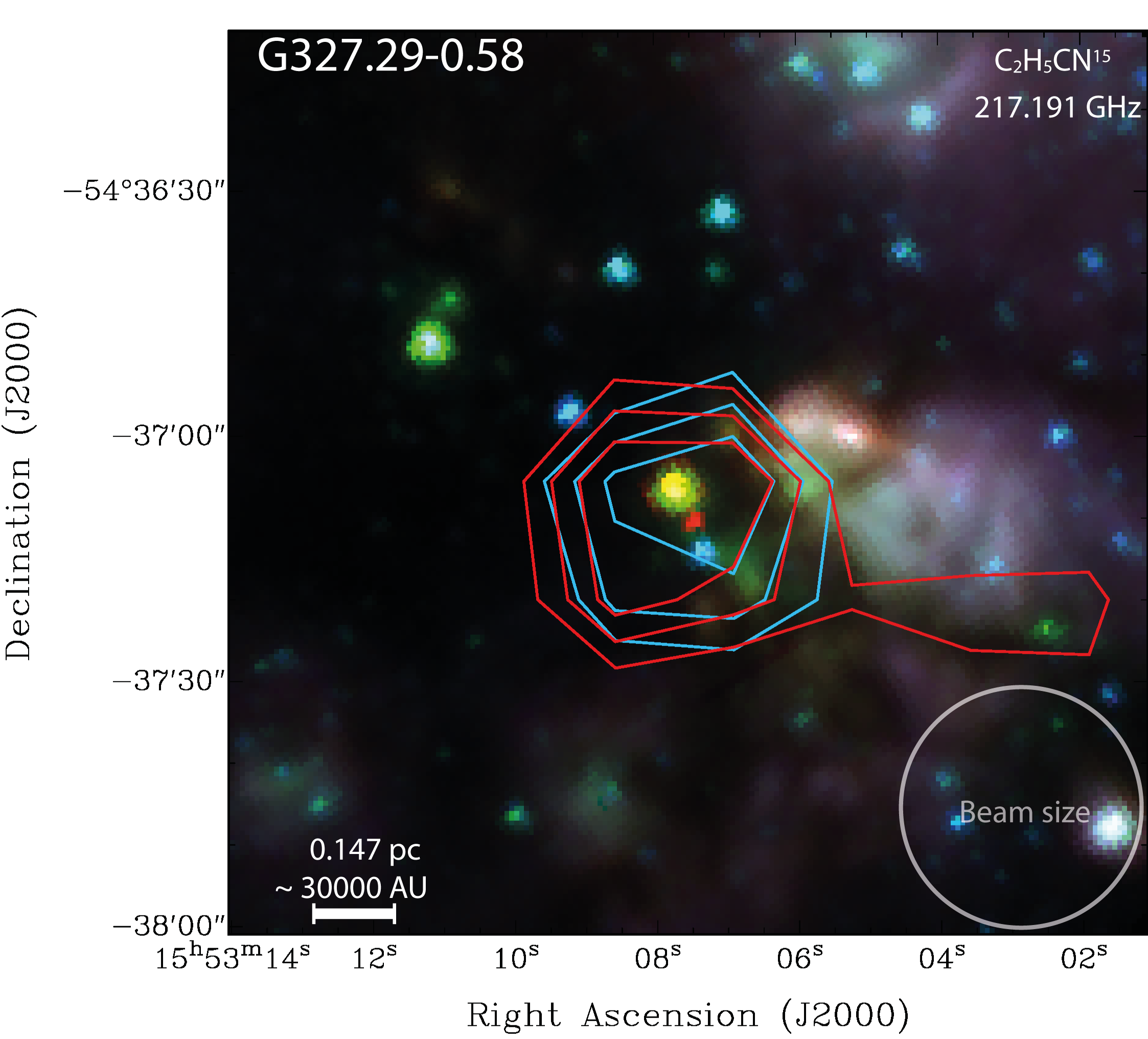}
	\caption{Contour map of the blue- and red-shifted emission of the blended line conformed mainly by ethyl cyanide and acetone at 217.191 GHz towards the G327.29-0.58 region. The emission is tracing the core region. The contours show the emission up from 66\%+10\% peak emission.  The shifted emissions were integrated outward from the central 8 km s$^{-1}$ to reach the FWZP for each emission line.}
	\label{ethyl_cyanide}
\end{figure}

The blue- and red-shifted components of the 217.191 GHz transition overlap towards the central region in a spherical morphology, which relates this iCOM to the central massive protostar(see Figure \ref{ethyl_cyanide}). 

\subsection{Vynil Cyanide}
We have two sources that show a contribution of C$_2$H$_3$CN: G327.29-0.58 and G351.77-0.54. In both cases the 216.936 GHz line is the only transition within our bandwidth, being this identification tentative. Nevertheless, this molecule has been already reported towards massive protostars \citep{Codella2013A&A...550A..81C,Fu2016RAA....16..182F}, and is therefore expected. The emission line profile in both sources is narrow and their contour maps are concentrated to the central massive protostar and exhibits a circular morphology for $\sim 40''$. Their column density is about one order of magnitude lower than the O-bearing iCOMs that have been also reported in other investigations \citep{Csengeri2019A&A}.

\subsection{Methylenecyclopropene}
The proposed transition of methylenecyclopropene ($c-$C$_3$H$_2$)CH$_2$  at 217.428 GHz was only observed towards the most chemically rich source, G327.29-0.58. Since we identified just one transition, this reduces its reliability. The observed emission line is close to 217.428 GHz, and therefore could be related to the more common isotopologue of $^{13}$CN, but this species was discarded as its fit implies the existence of other higher intensity transitions which should appear in our bandwidth. Note that this iCOM has not been detected in the ISM\footnote{Set of detected molecules in space according to CDMS https://cdms.astro.uni-koeln.de/classic/molecules} but it is already listed in CDMS \citep{Muller2005JMoSt.742..215M}.


\section{Summary and conclusions}
We have reported the detection and identification of six iCOMs towards eleven HMSFRs selected from the ATLASGAL and EGOs catalogs, previously related to outflow activity,  enlarging the existing outflow sample associated with iCOMs \citep{vandis93,Leurini13,garay2007,palau2017,leflo2017}. The identified iCOMs are methanol (CH$_3$OH), methyl formate (CH$_3$OCHO), acetone (CH$_3$COCH$_3$), vinyl cyanide (C$_2$H$_3$CN) and ethyl cyanide (C$_2$H$_5$C$^{15}$N). 

Our identification of iCOMs has been tested by LTE modeling using the XCLASS software, from which we obtained the physical parameters N$_{tot}$, $T_{kin}$, $v_{width}$, and $v_{offset}$. The agreement between the modeled synthetic spectra and the observed spectra was evaluated by a classical $\chi^{2}$ test. This test showed a statistically significant concordance as the critical $\chi^{2}_{crit}$ for a 0.05 p-value was under-passed ($\chi^{2}_{crit}$ < $\chi^{2}_{obs}$) in all but one of our sources (G340.97-1.02). 

The observed high abundance of O-bearing molecules points out the expected desorption from the ice mantles due to outflow-induced shocks in these HMSFRs. This premise was evaluated using the ubiquitously observed iCOM transition of methanol at 216.945 GHz. For this, we tested the correlation between the integrated emission of this line versus the integrated emission of the shock tracer SiO at 217.105 GHz. The correlation has shown an increasing linear behavior with a good significance level, which reinforces the premise that shocks could release the iCOMs trapped into the dust and mantles. This desorption could help the proliferation of second-generation iCOMs, since, as mother precursor, methanol has proved to trigger the production of other CH$_3$-bearing molecules \citep{Balucani2015}, such as CH$_3$OCHO and CH$_3$COCH$_3$ that were observed in our sample.

The proportional relation between methanol and SiO has been observed in the collection of our spectra (Figure \ref{xclass_all_spec1}). The peak of the methanol line is observed to be smaller than the SiO one, in all but G327.29-0.58, and the integrated line area is always larger for SiO than methanol. This relation could be a key ratio to describe the energetic process which are taking place in this outflow stage of HMSFRs. 

Other N-bearing species such as vinyl cyanide (C$_2$H$_3$CN) and an ethyl cyanide isotopologue (C$_2$H$_5$C$^{15}$N), have been observed only in three objects: G327.29-0.58, G351.77-0.54, and G034.41+0.24. The identification of vinyl cyanide is not unambiguously confirmed as it only shows one transition in our bandwidth. The column densities of N-bearing species are one to two orders of magnitude lower than the O-bearing CH$_3$OH and CH$_3$OCHO. 
The column density of vinyl cyanide goes from \num{3.7E14} to \num{1.7E15} and the ethyl cyanide goes from \num{8.2E14} to \num{6.4E15}. Whereas the column density among methanol goes from \num{2.4E15} to \num{4.7E17} and the MF goes from \num{5.7E13} to \num{3.6E16}. 

Acetone has been confirmed through four emission lines towards G034.26+015 and G327.29-0.58, with column density values of \num{1.5e13} and \num{1.7e15}, respectively, its abundances are one to two orders of magnitude lower than methanol.

We have proposed a faint line emission produced by methylenecyclopropene. This emission line is observed towards G327.29-0.58 at 217.428 GHz, and could not be modeled by other simpler molecules, since they would produce other high-intensity lines within our bandwidth that we did not detect. 



We made maps of the blue- and red-shifted emission of the widest  iCOMs lines. In general, this emission is mainly tracing the central core of the massive protostars. Nevertheless, some of the sources have shown an iCOM extended emission with offset peak positions of the blue- and red-shifted emission, covering from  $20000$ to $100000$ AU; hence, this could be related to movements external to the compact core, such as large scale low-velocity outflows. Unfortunately, due to our angular resolution, these outflows are strongly affected by beam dilution.

In this context, our sample provides an excellent set of candidates to carry out further studies towards the high mass outflows regime. Observations at high angular resolution are highly desirable to avoid the beam dilution of these faint iCOMs features. Since iCOMs have proven to better trace the compact cores of HMSFRs, high angular resolution observations could give hints for the innermost mechanisms at work in the formation of massive stars \citep[as in][]{palau2017,Csengeri2019A&A}.

Further studies with a wider bandwidth coverage 
are desirable to observe more transitions, 
and with these better constrain the modeling parameters of the identified iCOMs. 

A bigger sample is also convenient to further constrain the correlation of methanol towards SiO shocks in HMSFRs.

\textbf{Acknowledgements.-} We thank the support of researchers from the National Institute of Astrophysics, Optics and Electronics (INAOE), the Large Millimeter Telescope (LMT) and the Institute of Radio Astronomy and Astrophysics (IRyA) during the development of this investigation. We are also grateful to the Mexican funding agency Consejo Nacional de Ciencia y Tecnología (CONACYT) for its support to the INAOE's graduate program in Astrophysics. A.P. acknowledges financial support from the UNAM-PAPIIT IN111421 grant, the Sistema Nacional de Investigadores of CONACyT, and from the CONACyT project number 86372 of the `Ciencia de Frontera 2019’ program, entitled `Citlalc\'oatl: A multiscale study at the new frontier of the formation and early evolution of stars and planetary systems’, México. \clearpage





\appendix



\section{XCLASS spectra}\label{sec:xclass_spec}
In the following, we present the corresponding spectra towards all our sample and their corresponding XCLASS best fit. The physical parameters to get the synthesized spectra are shown in table \ref{xclass_result}.

\begin{figure}[htpb!]
	\centering  
	\includegraphics[width=0.9\linewidth,height=17.8cm]{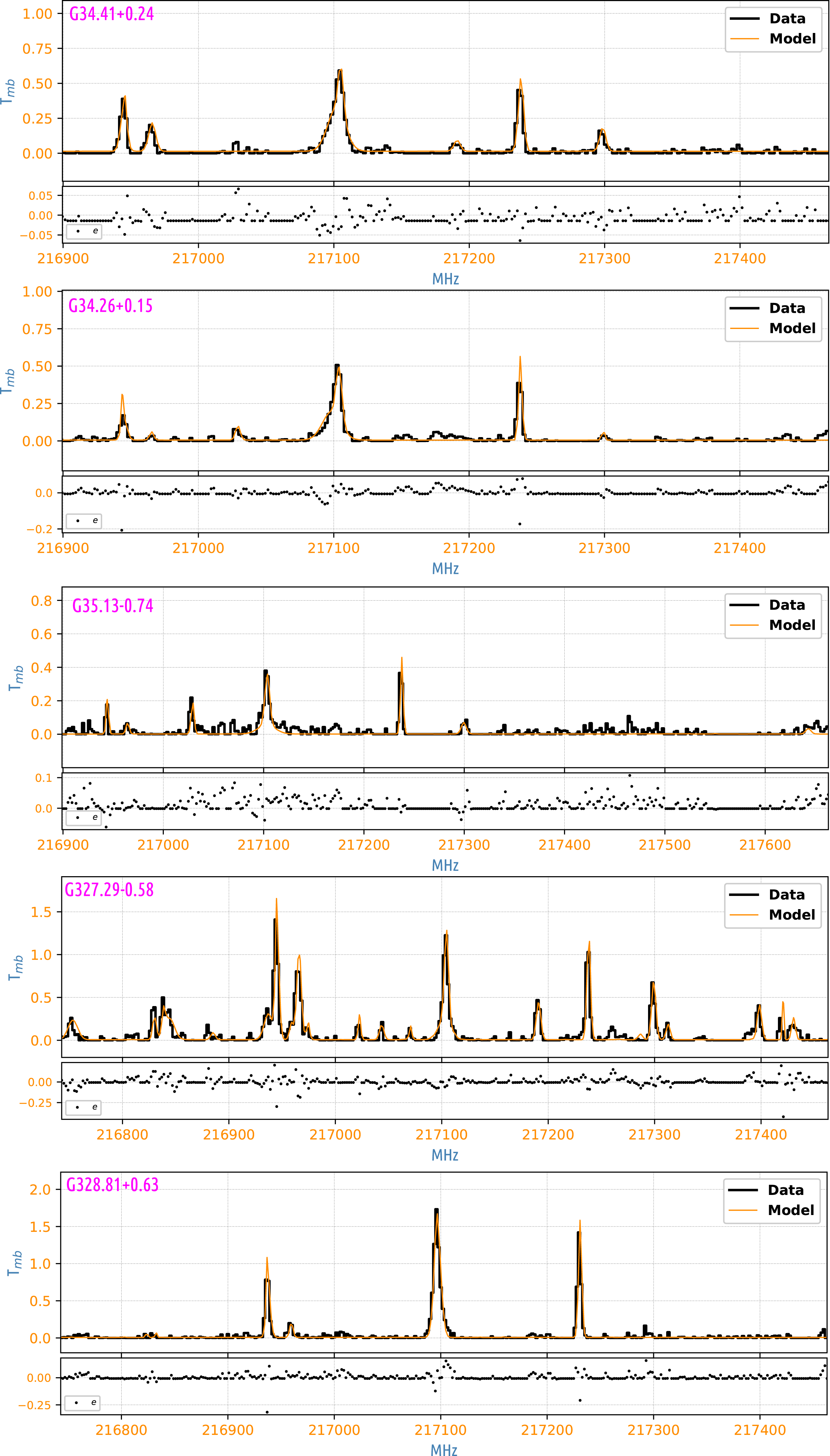}
	\caption{Collection of the spectra over our entire sample and their corresponding synthetic model calculated using XCLASS. The \textit{Black} solid line shows the observed spectra, the \textit{Orange} solid line shows the fit obtained using the XCLASS software.  In the lower subplot, we plotted the dispersion of the residuals $(e=I_{obs}-I_{model})$.}
	\label{xclass_all_spec1}
\end{figure}
\begin{figure}[htpb!]
	\ContinuedFloat
\centering 
\includegraphics[width=0.9\linewidth,height=21.5cm]{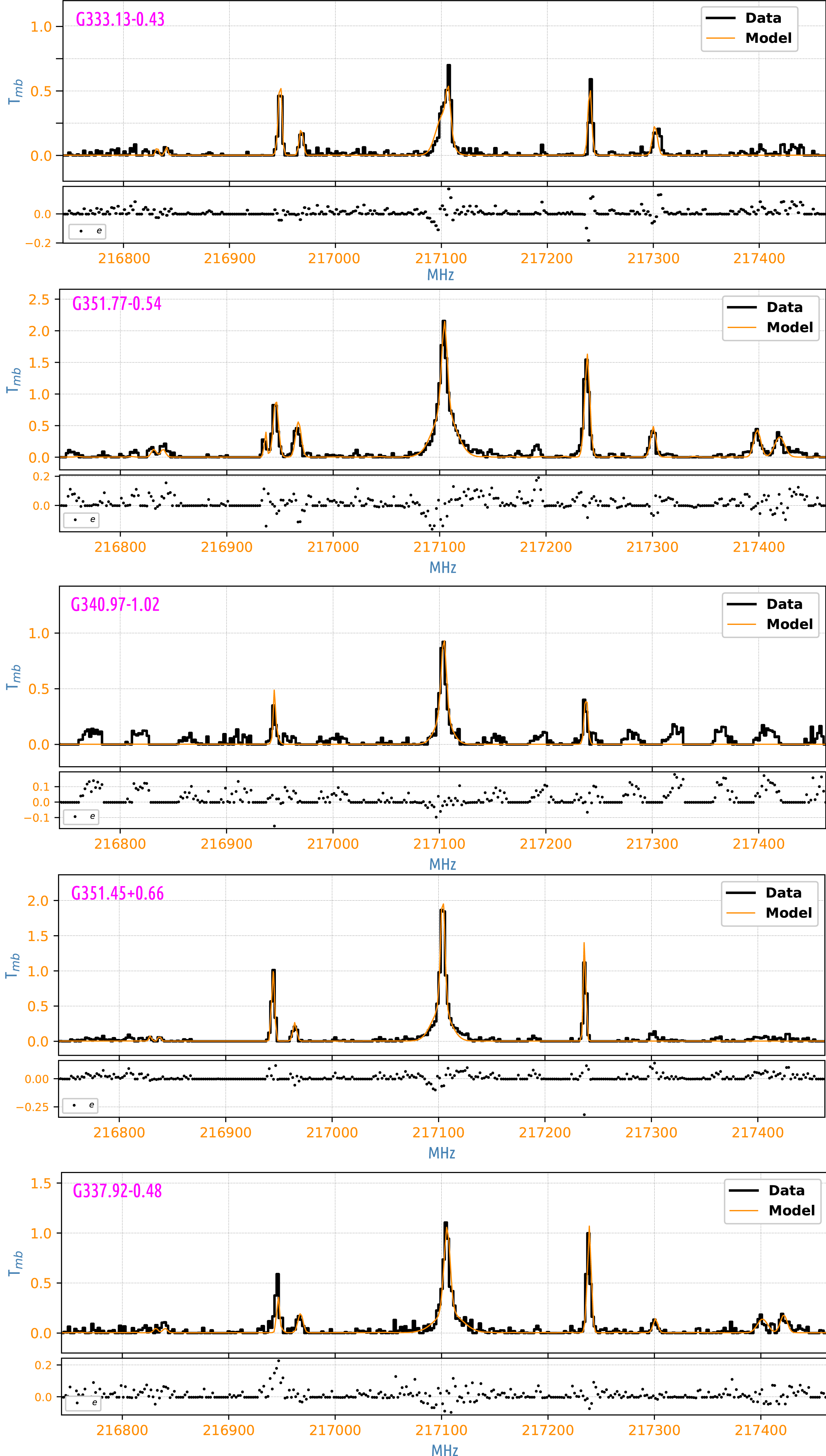}
	\caption{Continued.}
\end{figure}
\clearpage

\section{Integrated maps}
In the following, we present the complete integrated emission maps over each source at which we have been detected wider iCOMs lines. The images consist of a three-band composition of the IRAC camera of 3.6, 4.5, and 8.0$\mu$m bands, colored code as Blue, Green, and Red, respectively \citep{Cyganowski2008} and an over-plotted contours of the integrated emission of the iCOMs. The contours goes from 66 to 90\% of the peak emission with a spacing of 10\%. The shifted emission was integrated avoiding the central 8 kms$^{-1}$ until the emission reach their FWZP.
Considering that our sample has an average distance of 3.7 kpc, 10$''$ would correspond to 0.18 pc 
(a more precise physical scale is labeled for each source).

\begin{figure}[htpb!]
	\centering 
	\includegraphics[width=0.8\linewidth,height=18.1cm]{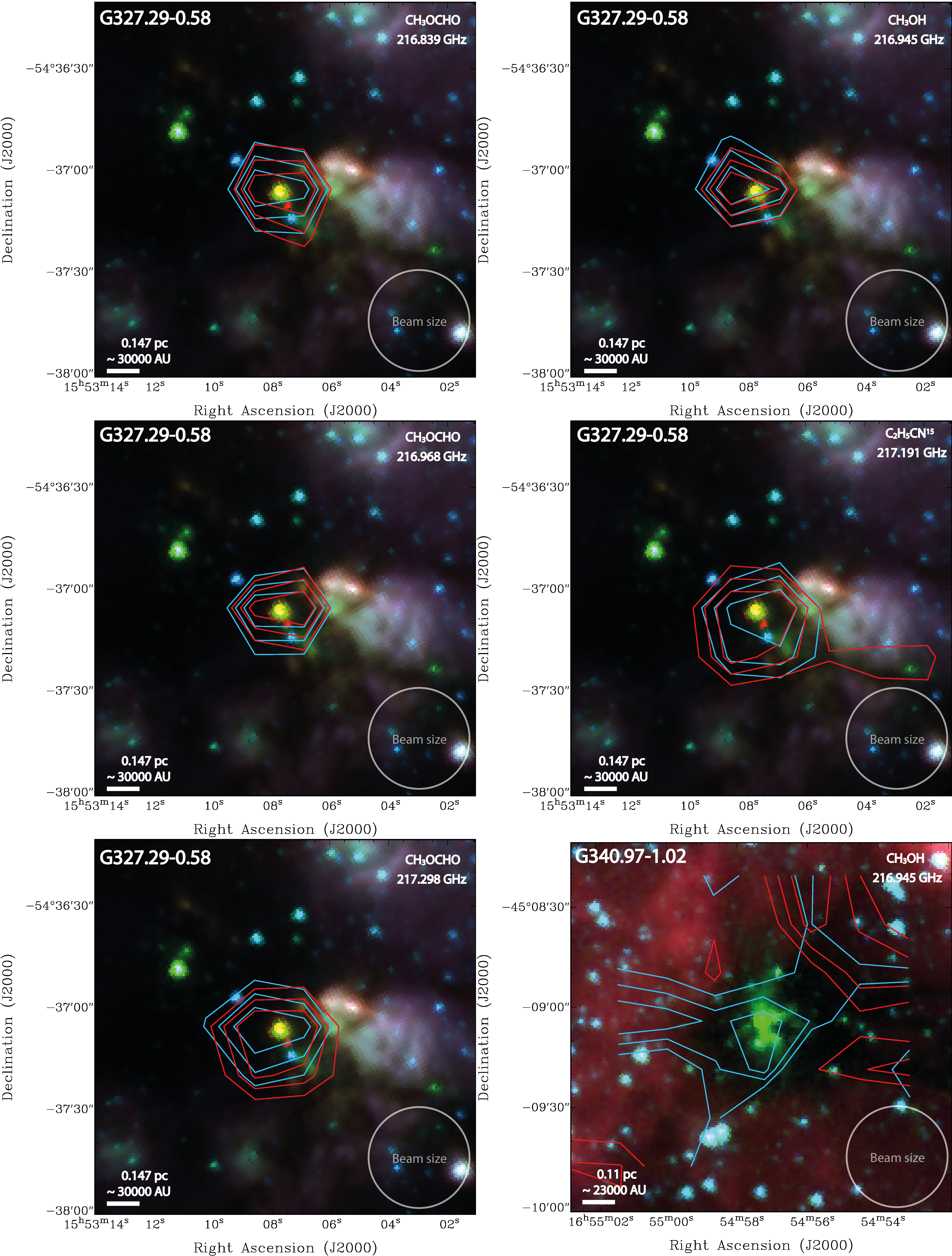}
	\caption{Integrated maps for the high intensity iCOMs lines on the different transitions.}\label{Maps}
\end{figure}

\begin{figure}[htpb!]
	\ContinuedFloat
\centering 
\includegraphics[width=0.95\linewidth]{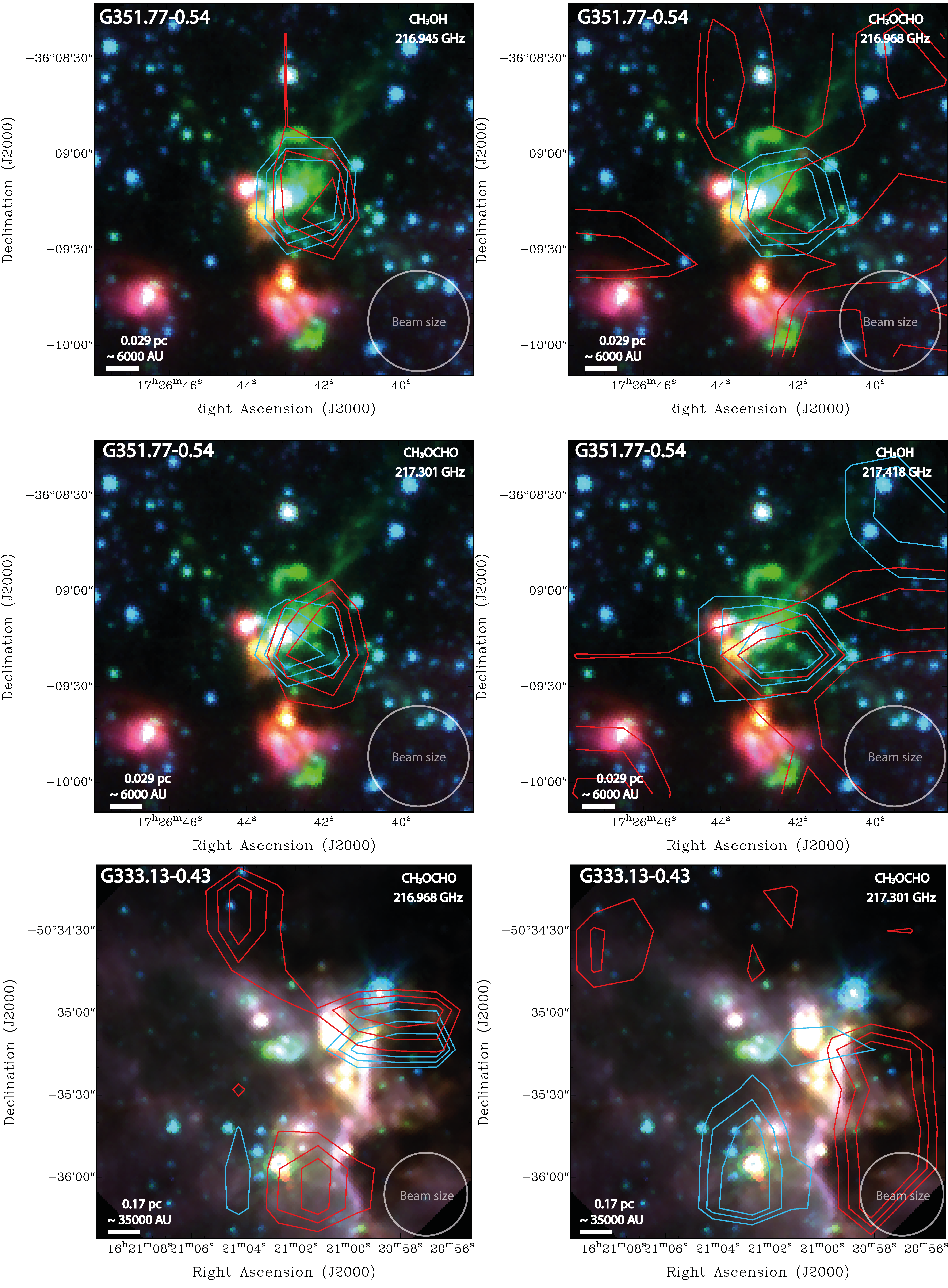}
	\caption{Continued.}
\end{figure}

\begin{figure}[htpb!]
	\ContinuedFloat
	\centering 
\includegraphics[width=0.95\linewidth]{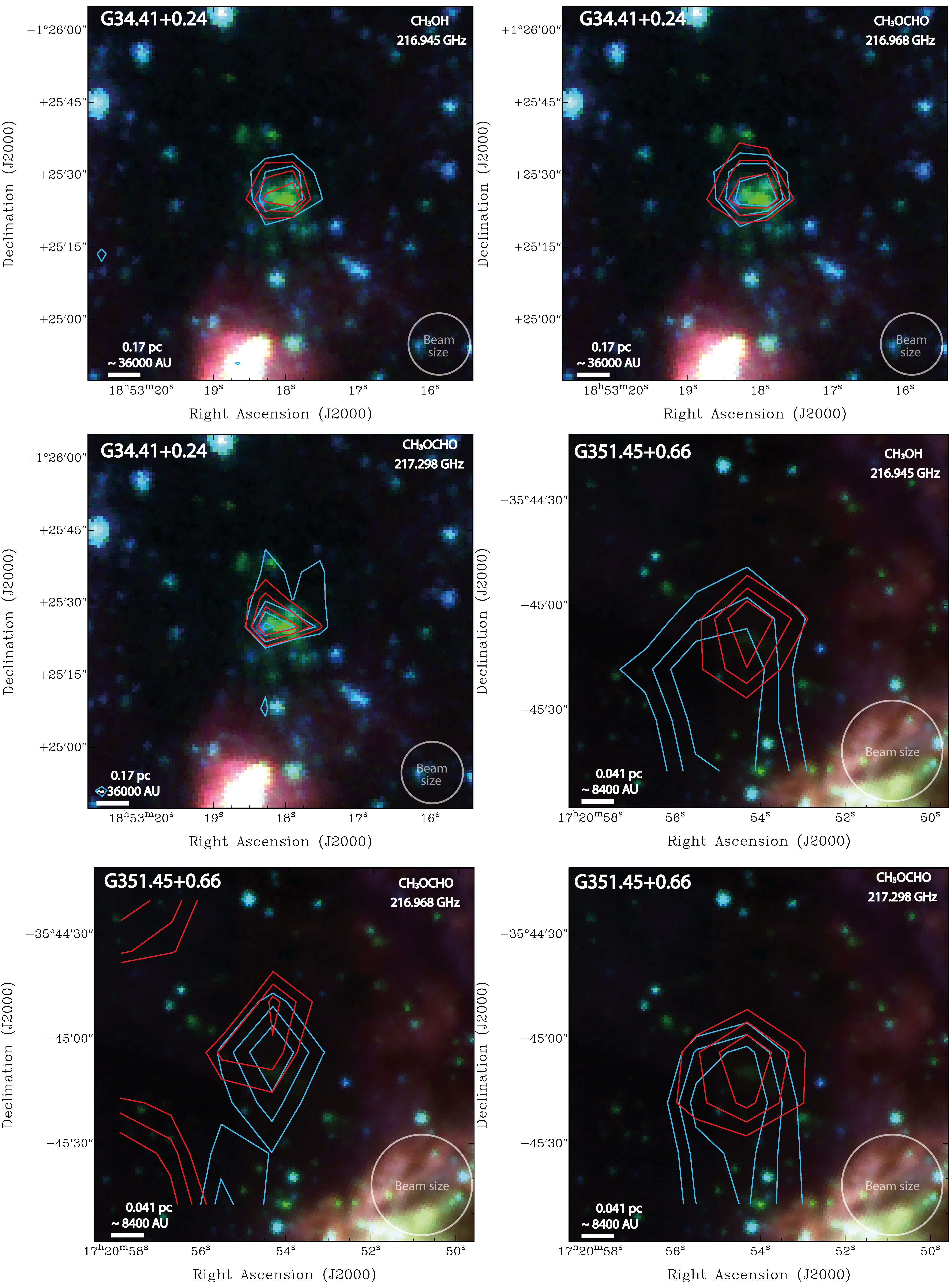}
	\caption{Continued.}
\end{figure}

\begin{figure}[htpb!]
	\ContinuedFloat
	\centering 
\includegraphics[width=0.51\linewidth]{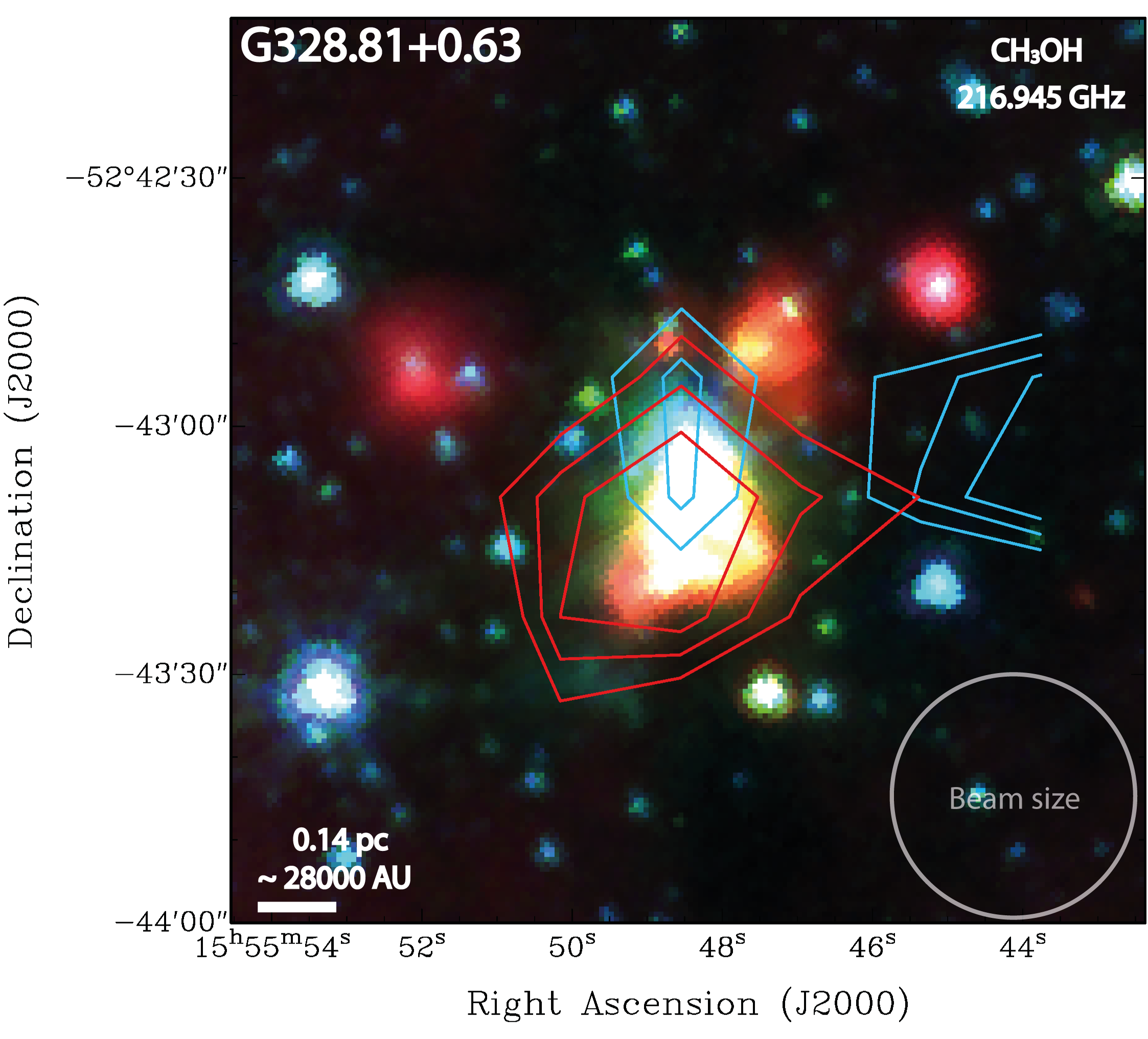}
	\caption{Continued.}
\end{figure}

\clearpage
\bibliographystyle{aasjournal}
\bibliography{library}

\end{document}